\documentclass[%
pre,
tightenlines,
showpacs,showkeys,
a4paper,12pt
]{revtex4}
\usepackage{dcolumn}
\usepackage{amssymb,amsmath,stmaryrd,array}
\usepackage{graphicx}
\usepackage{subfigure}

\renewcommand{\thesubfigure}{(\alph{subfigure})}

\makeatletter
\renewcommand{\@thesubfigure}{\thesubfigure\space}

\newcommand{\vc}[1]{\mathbf{#1}}
\newcommand{\uvc}[1]{\mathbf{\hat #1}}

\newcommand{\triang}{\mathrm{triang}}
\newcommand{\squar}{\mathrm{squar}}
\newcommand{\sine}{\mathrm{sine}}

\newcommand{\dd}{\mathrm{d}}

\newcommand{\spn}{\mathrm{S}}
\newcommand{\anc}{\mathrm{anch}}
\newcommand{\eff}{\mathrm{eff}}

\newcommand{\mx}{\mathrm{max}}
\newcommand{\mn}{\mathrm{min}}
\newcommand{\st}{\mathrm{st}}
\newcommand{\shf}{\mathrm{sh}}

\newcommand{\FLC}{\mathrm{FLC}}
\newcommand{\chiC}{C$^\ast$}
\newcommand{\chiA}{A$^\ast$}

\newcommand{\myarrow}[1]{\ensuremath{\xrightarrow{{#1}^\circ\mathrm{C}}}}

\newcommand{\sca}[2]{\bigl({#1}\cdot{#2}\bigr)}
\newcommand{\avr}[1]{\left\langle{#1}\right\rangle}

\newcommand{\pdr}[2]{\frac{\partial #1}{\partial #2}}

\newcommand{\degc}{$^\circ$C}
\newcommand{\mum}{$\mu$m}

\newcommand{\sign}{\mathop{\rm sign}\nolimits}

\makeatother

\begin{document}
\DeclareGraphicsExtensions{.jpg,.pdf}
\title{Switching dynamics of surface stabilized ferroelectric
liquid crystal cells: effects of anchoring energy asymmetry}

\author{Alexei~D.~Kiselev}
\affiliation{%
 Institute of Physics of National Academy of Sciences of Ukraine,
 prospekt Nauki 46,
 03028 Ky\"{\i}v, Ukraine} 

\email[Email address: ]{kiselev@iop.kiev.ua}

 \author{Vladimir~G.~Chigrinov}
\affiliation{%
 Hong Kong University of Science and Technology,
 Clear Water Bay, Kowloon, Hong Kong
 }

 \email[Email address: ]{eechigr@ust.hk}

\author{Eugene~P.~Pozhidaev}
 \affiliation{%
P.N. Lebedev Physics Institute of Russian Academy of Sciences,
Leninsky prospect 53, 117924 Moscow, Russia
 }

\date{\today}

\begin{abstract}
We study both  theoretically and experimentally
switching dynamics in surface stabilized
ferroelectric liquid crystal (SSFLC) 
cells with asymmetric boundary conditions.
In these cells the bounding surfaces are treated differently
to produce asymmetry in their anchoring properties.
Our electro-optic measurements of  
the switching voltage thresholds, $V_{+}$ and $-V_{-}$, 
that are determined by the peaks of the reversal polarization current 
reveal the frequency dependent shift of the hysteresis loop, $V_{+} - V_{-}$. 
We examine the predictions of the uniform dynamical model
with the anchoring energy taken into account.
It is found that the asymmetry effects are dominated 
by the polar contribution to the anchoring energy. 
Frequency dependence of the voltage thresholds
is studied by analyzing the properties of
time-periodic solutions to the dynamical equation
(cycles). For this purpose, we apply the method that uses the parameterized 
half-period mappings for the approximate model
and relate the cycles to the fixed points of the composition of two
half-period mappings.
The cycles are found to be unstable and
can only be formed when the driving frequency is lower
than its critical value.
The polar anchoring parameter is estimated by making a comparison
between the results of modelling and the experimental data for
the shift vs frequency curve. 
For a double-well potential considered as a deformation of the
Rapini-Papoular potential, the branch of stable cycles emerges
in the low frequency region separated by the gap 
from the high frequency interval for unstable cycles. 
\end{abstract}

 
\maketitle

%%%%%%%%%%%%%%%%%
\section{Introduction}
\label{sec:intro}
%%%%%%%%%%%%%%%%%

As it was shown by Meyer \textit{at al.} in 1975~\cite{Meyer:jpfl:1975},
the origin of ferroelectric ordering
in smectic-\chiC\ (Sm-\chiC) liquid crystals
with the spontaneous polarization
normal to the tilt plane
is closely related to the reduction of symmetry  
caused by chirality of the tilted Sm-\chiC\ phase. 
This phase is thus naturally ferroelectric
and the Sm-\chiC\ liquid crystals are also known as the
\textit{ferroelectric liquid crystals} (FLC)
(a detailed description of FLCs can be found 
e.g. in a collection of reviews~\cite{Taylor:collec:1991}
and in a more recent monograph~\cite{Musevic:bk:2000}).

Over the past three decades
FLCs have been attracted considerable attention
not only as a vital issue in the condensed matter
physics but also
as promising materials for applications in electro-optic switching
devices. The first such device
was due to Clark and Lagerwall~\cite{Clark:apl:1980}.
They studied electro-optic response of 
FLC cells confined between two parallel plates
subject to homogeneous boundary conditions
and made thin enough to suppress the bulk chiral Sm-\chiC\ 
helix. It was found that
such cells~---~the so-called
\textit{surface-stabilized ferroelectric liquid-crystal} (SSFLC)
cells~---~exhibit high-speed, bistable electro-optical switching
between orientational states stabilized by surface interactions.

The response of chiral Sm-\chiC\  liquid crystals
to an applied electric field
is characterized by fast switching times due to
linear coupling between the field and the spontaneous polarization.
There is also a threshold voltage necessary for switching to occur
and the process of bistable switching is typically accompanied
by a hysteresis.
The switching times and the threshold voltages
may considerably vary depending on
the wave form, the amplitude and the frequency
of applied (driving) voltages.

In early studies~\cite{Hands:apl:1982,Hands:prl:1983,Shill:crt:1986}, 
it was found that the competition between elastic,
electrostatic and surface energies may result in
different regimes of switching and field induced transitions
in FLC cells.
Certain regimes such as the high voltage regime
can be described using the theoretical approach
based on the assumption that director and polarization fields
are spatially homogeneous.
This approach provides uniform switching models
where the effects of electrostatic and surface interactions
are incorporated into an effective potential
governing the dynamics of switching.
Such models can be readily applied to interpret experimental data.

In particular, the uniform model of switching supplemented 
with an elastic-like term was examined in~\cite{Dahl:pra:1987}
as a useful tool to describe
experimental behavior of the polarization 
reversal current in SSFLC cells.
Similarly, the uniform theory was applied
to determine the rotational viscosity and 
the anchoring energy strength
from the experimental results on
the response time measured as a function of pulse voltage 
in SSFLC cell~\cite{Sako:apl:1997}.
The field-reversal method suggested in~\cite{Panov:lc:2001}
to measure the spontaneous polarization,
the switching time, the rotational viscosity
and the dc conductivity also relies on the uniform model.
(Recent discussions, applications and generalizations of uniform models
can be found e.g. 
in~\cite{Nakagawa:jjap:2000,Elston:lc:2001,Pauwels:lc:2001,Essid:lc:2005}.)

In this paper we  are aimed to study 
the effects of surface anchoring energy
in switching dynamics of asymmetric SSFLC cells
where  nonidentical aligning films
impose different boundary conditions at the substrates.
The effect of our particular concern is
the frequency dependent shift of the hysteresis loop
observed in our experiments.

In the theoretical part of the paper,
we adapt a systematic approach and
examine predictions of the uniform theory 
where the effects of asymmetry are caused
by the polar contribution to the anchoring energy potential.
The  voltage thresholds in relation to the driving frequency
are studied by using the method suggested to
explore the properties 
of periodic solutions to the dynamical equation.

The layout of the paper is as follows.
In Sec.~\ref{sec:model} 
we derive the effective potential and formulate the model.
The simplified case with the anchoring energy neglected
is discussed so as to clarify the assumptions
taken to obtain frequency dependent threshold voltages.

Analysis of the switching dynamics is performed 
in Sec.~\ref{sec:switch-dyn}. 
We present the approach that uses the parameterized 
half-period mappings for the approximate model
to study the time-periodic solutions of the dynamical equation.
These solutions are called the cycles and related to the fixed
points of the composition of the half-period mappings. 
The method is applied to the case of square wave voltages.
Analytical relations for the conditions of complete switching,
the switching times and
the critical frequency bounding the region of cycles from above 
are obtained. For sine-wave and triangular voltages,
we comment on the numerical procedure and show that,
qualitatively, the results remain unchanged.

Experimental details and the results
of electro-optical measurements for the switching voltage thresholds
are given in Sec.~\ref{sec:experim}.
The experimental data and the theoretical results of
Sec.~\ref{sec:switch-dyn} are used to model the process of switching
and to estimate the polar anchoring parameter.
Finally, in Sec.~\ref{sec:discussion} we present our results
and make some concluding remarks.
In the Appendix we extend our analysis to the dynamical model
with the double-well effective potential and study the branches
of stable and unstable cycles.

%%%%%%%%%%%%%%%%%%%%%%%%%%%%%%%%%%%%%%%%%%
\begin{figure*}[!tbh]
%\vskip5mm
\centering
\resizebox{150mm}{!}{\includegraphics*{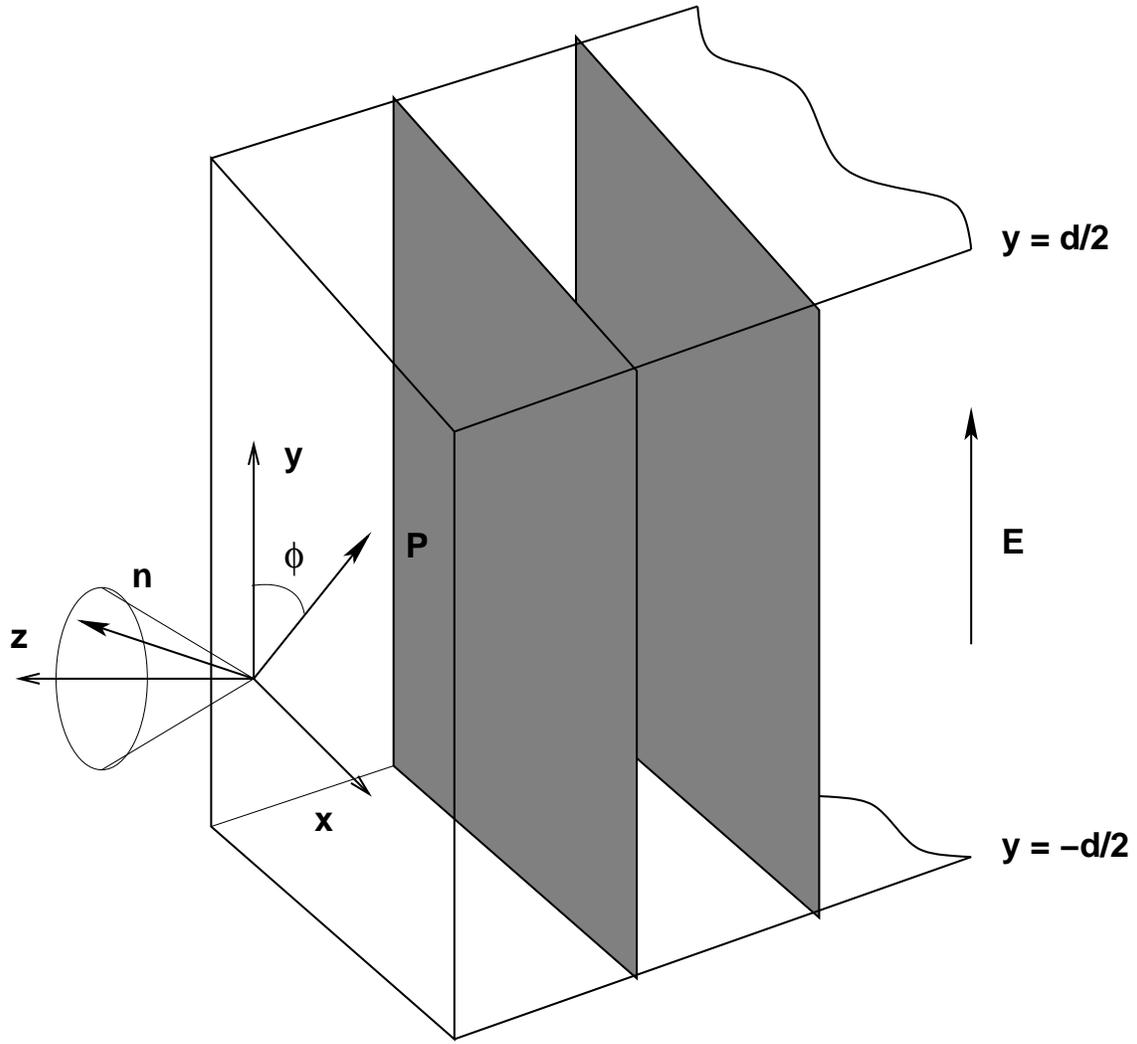}}
\caption{%
Schematic representation of bookshelf geometry
illustrating planar arrangement of smectic layers.
The polarization vector lies in the $x-y$ plane
and $\phi$ is the azimuthal angle around the smectic cone.
External electric field is directed along the $y$-axis
normal to the boundary surfaces, $y=-d/2$ and $y=d/2$.
}
\label{fig:bookshelf}
\end{figure*}

\begin{figure*}[!tbh]
%\vskip5mm
\centering
\resizebox{150mm}{!}{\includegraphics*{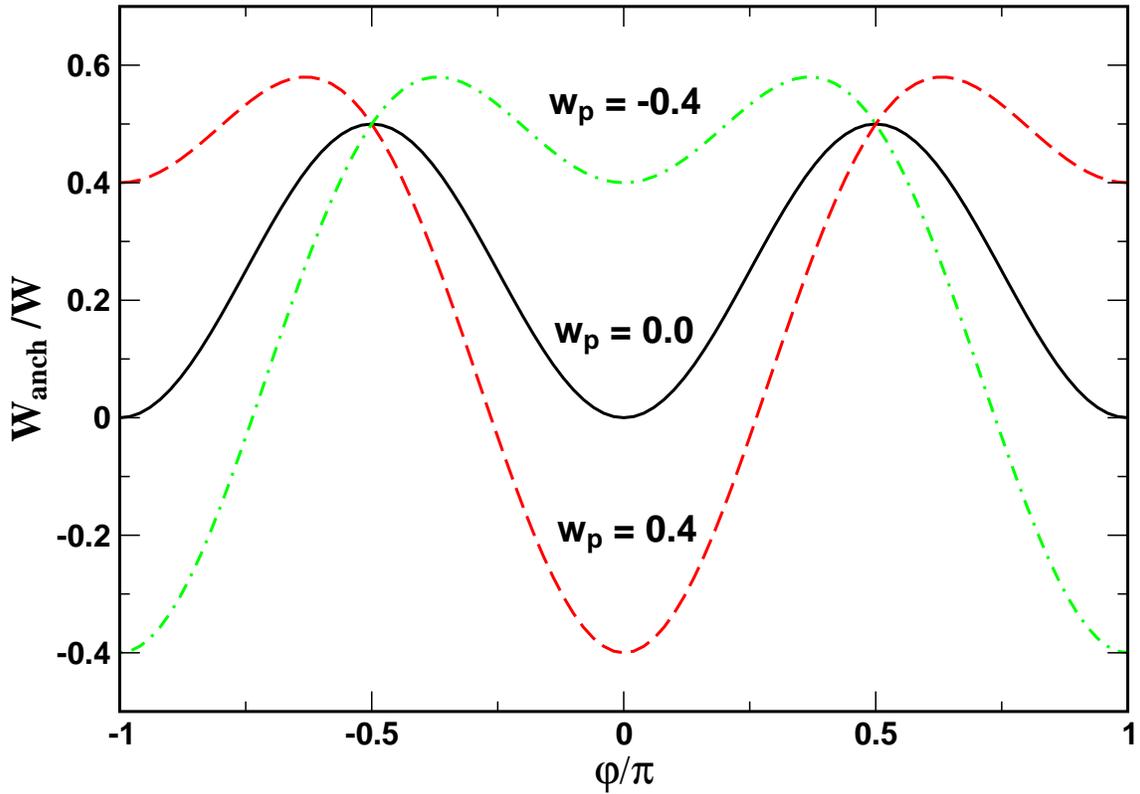}}
\caption{%
Anchoring energy as a function of the azimuthal angle $\phi$
at various values of the polar anchoring parameter $w_{p}\equiv W_{p}/W$.
}
\label{fig:poten-phi}
\end{figure*}

\begin{figure*}[!tbh]
%\vskip5mm
\centering
\resizebox{150mm}{!}{\includegraphics*{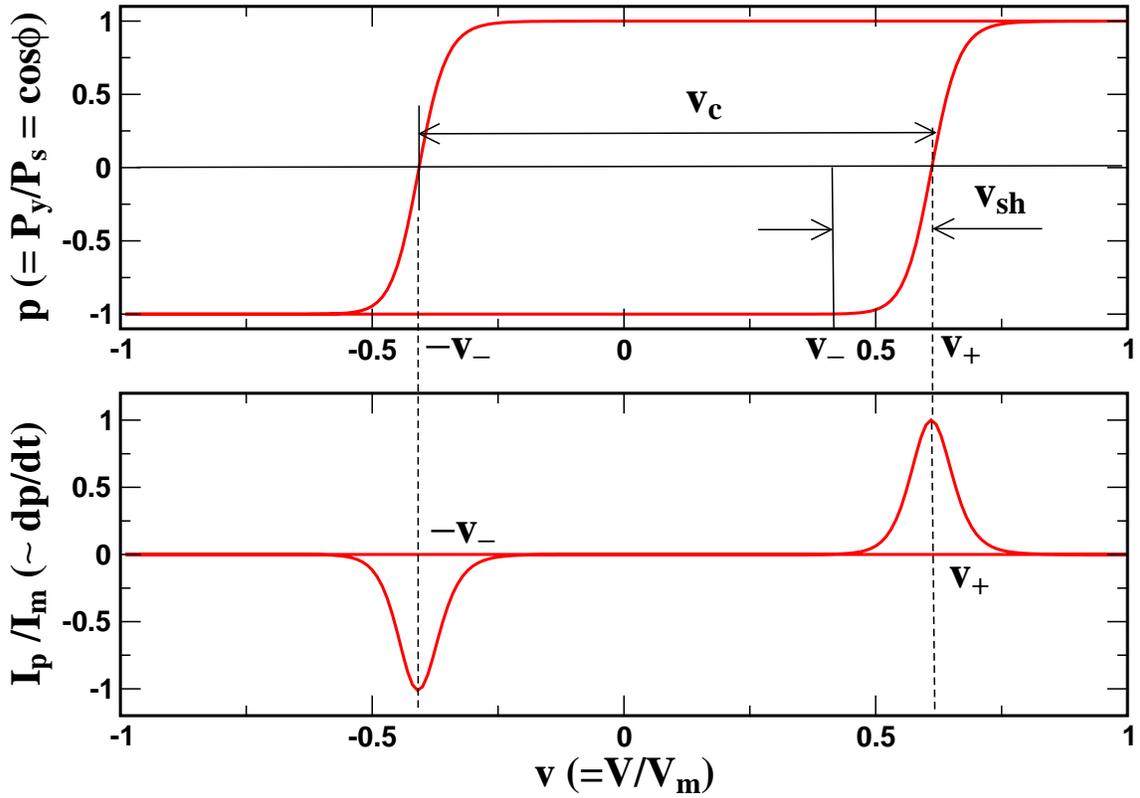}}
\caption{%
Hysteresis loop.
}
\label{fig:hysteresis}
\end{figure*}

\begin{figure*}[!tbh]
%\vskip5mm
\centering
\resizebox{155mm}{!}{\includegraphics*{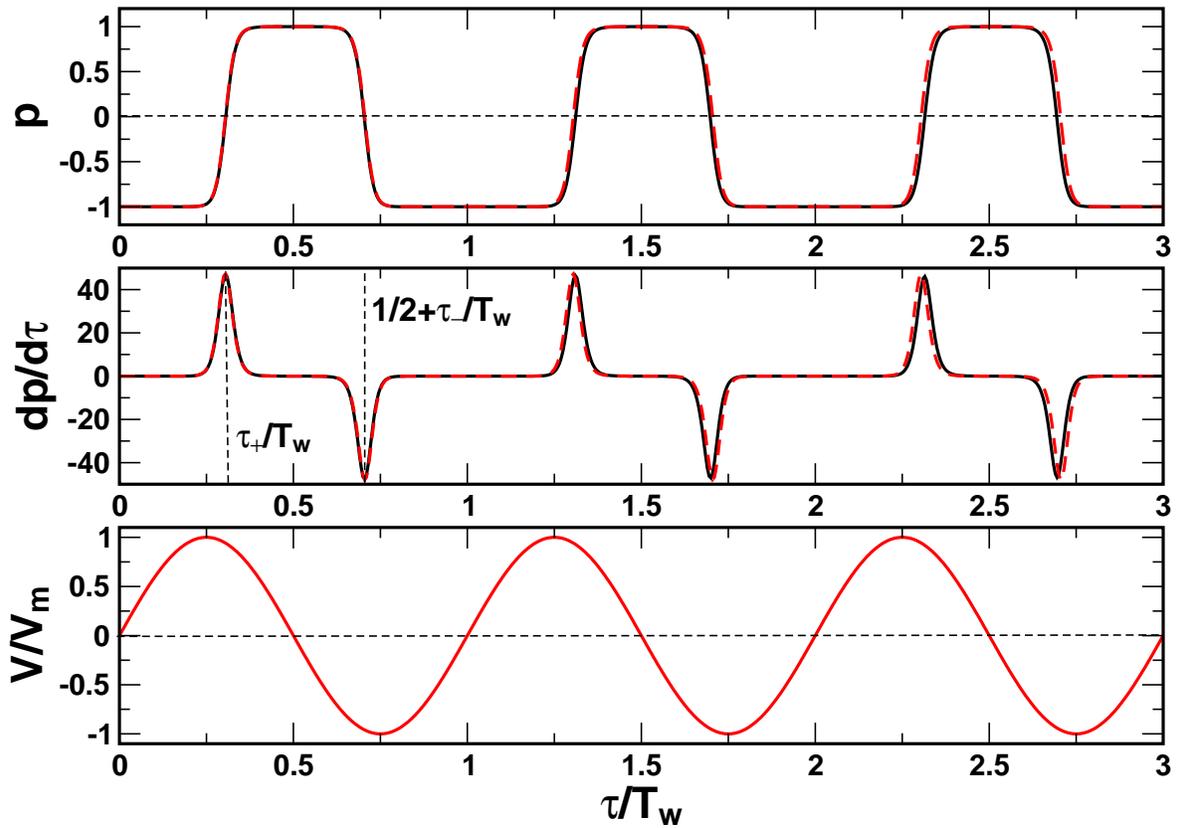}}
\caption{%
Normalized polarization and polarization reversal current evolving in time
under sine-wave driving voltage
at $w_p=0.2$, $T_{w}=0.8$ and $r_e=50$.
Solid lines represent the results obtained by solving
Eq.~\eqref{eq:dyn-phi} numerically.
Analytical results (dashed lines) are computed using the piecewise linear
approximation~\eqref{eq:tanh-approx}.
}
\label{fig:time-diagram}
\end{figure*}

%%%%%%%%%%%%%%%%%%
\section{Model}
\label{sec:model}
%%%%%%%%%%%%%%%%%%

In this section we 
introduce necessary notations and
describe the model
that takes into account the effects due to aligning films.
The equation of motion governing switching
dynamics in asymmetric SSFLC cells
are derived and
the asymmetry induced polar contribution to the anchoring energy
 is found to play an important part in the problem
Frequency dependence of the  switching
voltage thresholds is analyzed using a simplified model
as an example demonstrating the key points underlying 
subsequent theoretical considerations.

%%%%%%%%%%%%%%%%%%%%%%%%%%
\subsection{Free energy}
\label{subsec:energy}
%%%%%%%%%%%%%%%%%%%%%%%%%%

We consider the FLC cell with a planar arrangement of
planar smectic layers usually referred to as the \textit{bookshelf} geometry.
This geometry is schematically presented in Fig.~\ref{fig:bookshelf}.
It is characterized by the director field  
\begin{equation}
  \label{eq:director}
  \uvc{n}=\uvc{z}\,\cos\theta+\sin\theta\,
(\,\uvc{x}\,\cos\phi+\uvc{y}\,\sin\phi\,),\quad
\uvc{k}=\uvc{y},
\end{equation}
where $\theta$ is the molecular cone angle,
$\phi$ is the azimuthal angle around the smectic cone, 
$\uvc{k}$ is the outward (inward) normal to the upper (lower)
substrate of FLC cell, $y=d/2$ ($y=-d/2$), 
and $d$ is the cell thickness.

The vector of the polarization, $\vc{P}$, and
the electric field inside the cell, $\vc{E}$, are given by 
\begin{equation}
  \label{eq:P_s}
  \vc{P}= P_{\spn}\,\uvc{p},\quad
\uvc{p}=\uvc{y}\,\cos\phi-\uvc{x}\,\sin\phi,\quad
\vc{E}=E\,\uvc{y}, 
\end{equation}
where $P_{\spn}$ is the spontaneous ferroelectric polarization.
The external driving voltage
\begin{equation}
  \label{eq:V_drv}
  V(t)=V_{\mathrm m}\,v(t)
\end{equation}
is characterized by the amplitude $V_{\mathrm m}$,
the frequency $f=1/T$ ($T$ is the period)
and the $T$-periodic function of time $v(t)$ describing the waveform
of the applied voltage.

For sinusoidal (sine-wave) driving voltages, the waveform function $v(t)$
is given by
\begin{equation}
\label{eq:sin-wv}
  v_{\sine}(t)=\sin(\omega t),
\quad
\omega = 2\pi/T,
\end{equation}
whereas
$T$-periodic continuations of the functions
$v_{\triang}(t)$ and $v_{\squar}(t)$:
\begin{align}
&
\label{eq:triang-wv}
v_{\triang}(t)=
\begin{cases}
4 t/T, & 0\le t/T\le 1/4,\\
2- 4 t/T, & 1/4\le t/T\le 3/4,\\
4 t/T-4, & 3/4\le t/T\le 1,
\end{cases}
\\
&
\label{eq:squar-wv}
v_{\squar}(t)=
\begin{cases}
1, & 0\le t/T< 1/2,\\
-1, & 1/2\le t/T< 1,
\end{cases}
\end{align}
give the waveform functions for
the triangular and square-wave forms, 
respectively.  

Similar to nematic liquid crystals (NLC),
when a FLC layer is brought into contact with 
aligning substrates, the energy of the FLC molecules in the
interfacial layer and thus the surface tension 
will be orientationally
dependent.  
The anisotropic part of the surface tension~---~the
so-called \textit{anchoring energy}~---~gives rise to the phenomenon
known as \textit{anchoring}, i.e., surface induced 
alignment of the FLC director along the vector of preferential orientation
referred to as the \textit{easy axis}.

Phenomenologically, the expression for the anchoring energy
can be written as a linear combination of the invariants constructed
from the surface normal, $\uvc{k}$, the FLC director, $\uvc{n}$ and the
unit polarization vector $\uvc{p}$.
For FLC cells, this gives the anchoring energy 
expressed in terms of  the scalar products
$\sca{\uvc{n}}{\uvc{k}}$ and
$\sca{\uvc{p}}{\uvc{k}}$ as follows
\begin{equation}
  \label{eq:anchoring}
W_{\anc}=\sum_{\nu=\pm\,1}
\left[
\frac{W_{\nu}}{2}\,\sca{\uvc{n}}{\uvc{k}}^2+\nu
W_{\nu}^{(P)}\sca{\uvc{p}}{\uvc{k}}
\right]_{y=\nu d/2},
\end{equation}
where
$W_{+}$ ($W_{-}$) is the strength of nonpolar anchoring
at the upper (lower) substrate;
similarly, $W_{+}^{(P)}$ and $W_{-}^{(P)}$  
is the polar anchoring strength at the upper and lower substrate,
respectively.

The first term in square  brackets on the right hand side of Eq.~\eqref{eq:anchoring}
represents the Rapini-Papoular surface potential~\cite{Rap:1969}
preserving equivalence between $\uvc{n}$ and $-\uvc{n}$.
This equivalence, however, can be broken
due to effects of polar ordering in the interfacial
layer~\cite{Parsons:prl:1978,McMullen:pra:1988}. 

As opposed to the case of nematic liquid crystals,
where the effects of surface induced polarity are mainly caused by
the quadrupole-dipole
interaction~\cite{Sluck:jpf:1993,Sluck:pre:1997},
one might expect that in FLCs the dominating factor
is the electrostatic interaction of
the spontaneous polarization $\vc{P}$ with the surface charges or
dipoles.
So, the polar anchoring term
on the right hand side of Eq.~\eqref{eq:anchoring}
is taken to be proportional to
$\sca{\uvc{p}}{\vc{k}}$.
This term 
introduces dependence on the polarity of the
polarization resulting from 
the ferroelectric polar surface interaction.

When the director field~\eqref{eq:director}
is spatially homogeneous,
the anchoring energy~\eqref{eq:anchoring} takes the simplified form
\begin{equation}
  \label{eq:anchoring-uni}
W_{\anc}=
\frac{W}{2}\,\sin^2\phi-
W_{P}\,\cos\phi,
\end{equation}
where
$W=(W_{-}+W_{+})\sin^2\theta$
and $W_{P}=W_{-}^{(P)}-W_{+}^{(P)}$.
From Eq.~\eqref{eq:anchoring-uni}
it is seen that, for symmetric cells with
$W_{-}^{(P)}=W_{+}^{(P)}$,
the polar terms add to zero.
In asymmetric FLC cells, the \textit{polar anchoring parameter},
$w_p=W_{P}/W$, is generally non-vanishing.

In Figure~\ref{fig:poten-phi}, the
potential~\eqref{eq:anchoring-uni} is plotted in relation to the azimuthal
angle $\phi$. 
It is illustrated that there are two local minima:
$\phi=0$ 
(the up state with $p\equiv\sca{\uvc{p}}{\vc{k}}=\cos\phi=+1$) 
and $\phi=\pi$
(the down state with $p=-1$).
When $w_p\ne 0$, the minima are not energetically equivalent
and one of the minima represents the metastable state. 
In addition, it is not difficult to obtain the relation
\begin{equation}
  \label{eq:meta-stability}
  |w_p|=|W_{P}/W| < 1
\end{equation}
as  the condition for the configurations with
$\phi=0$ and $\phi=\pi$ to be metastable.
If the metastability condition~\eqref{eq:meta-stability}
is broken the anchoring potential has only one minimum.
Similar results were previously reported in 
Refs.~\cite{Pauwels:lc:2001,Copic:pre:2003} and
the effects of asymmetry induced by the polar anchoring in
switching dynamics will be of our primary concern.
Throughout the paper 
we shall assume that, in the zero-field case, 
the up and down states are both metastable
and the condition~\eqref{eq:meta-stability} is satisfied.

In general, the polar anchoring energy
is known to be of considerable importance in the understanding of
static and dynamical behavior of chiral smectics in confined geometries.
For example, 
it may lead to the discontinuous Fr\'{e}edericksz 
transitions~\cite{Hands:prl:1983}
and is found to give rise to 
the surface electroclinic effect~\cite{Xue:prl:1990,Clark:lc:2001}. 
The magnitude of the polar anchoring strength
was also estimated from the experimental data
on the static dielectric susceptibility, the voltage coercivity
and the relaxation time~\cite{Pozhid:mm:1993}.

In a more recent
theoretical investigation~\cite{Mottram:pre:2000} 
into 
the effects of polar anchoring,
which induces ferroelectric ordering
close to the bounding surfaces 
of an antiferroelectric liquid crystal cell,
it was found that such ordering may result in
a coexistence of multiple zero-voltage ground states.

As far as the SSFLC cells are concerned,
the polar part of the anchoring energy
was considered theoretically in 
the papers~\cite{Copic:pre:2002,Copic:pre:2003,Callaghan:pre:2003}
dealing with
the so-called 
``thresholdless'' or ``V-shaped'' switching mode
of the optical response to an applied voltage.
It was found that monostable structures
exhibiting the V-shaped switching
can be formed as a result of
the polarization charge self-interaction accompanied by the effects
of the insulating alignment layers.

To see this, we briefly discuss a simple double-layer model. 
The starting point is the standard expression for the electrostatic part
of the FLC free energy per unit area
\begin{equation}
\label{eq:f_E}
F_{E}/A=\int_{-d/2}^{d/2} f_{E}\,\dd y,\quad
f_{E}= - \varepsilon_0\varepsilon_{yy} E^{\,2}/2 - P_{y} E,
\end{equation}
where 
$A$ is the area of the substrates,
$\varepsilon_0=8.854\times 10^{-12}$~Fm$^{-1}$
is the permittivity of free space and
$\varepsilon_{ij}=\varepsilon_{\perp} \delta_{ij}+
(\varepsilon_{\parallel}-\varepsilon_{\perp})n_{i} n_{j}$ 
is the FLC dielectric tensor.

For FLC sandwiched between two
insulating layers, we neglect the conductivity of FLC
and use
the boundary conditions for the normal component of 
the electric displacement field, $D_{y}$,
 to obtain the relations
\begin{equation}
  \label{eq:D_y}
  \varepsilon_0\varepsilon_{yy} E +P_{y}=
 \varepsilon_0\varepsilon_{1} E_{1}=
\varepsilon_0\varepsilon_{2} E_{2}
=D_{y},
\end{equation}
where $\varepsilon_{1}$ ($\varepsilon_{2}$) is the dielectric constant
of the lower (upper) insulating layer
and $E_{i}$ is the electric field inside the layers.
The voltage applied across the cell is given by
\begin{equation}
  \label{eq:V-E}
  V\equiv V(-d/2-d_1)-V(d/2+d_2)=
E_{1} d_{1}+E_{2} d_{2} + \int_{-d/2}^{d/2} E\,\dd y,
\end{equation}
where $d_{i}$ is the thickness of the layers.

We can now use Eq.~\eqref{eq:D_y}
to eliminate the electric fields, $E_{1}$ and $E_{2}$,
from the expression~\eqref{eq:V-E} and relate
the electric displacement field to the external electric field, $E_{0}$,
as follows 
\begin{equation}
  \label{eq:E_0-D}
  V/d\equiv E_{0}=
\eta D_{y}/\varepsilon_{0}+\avr{E},
\quad
\eta=(d_1/\varepsilon_1+d_2/\varepsilon_2)/d,
\end{equation}
where $\avr{\dots}=d^{-1}\int_{-d/2}^{d/2}\dots\dd y$.
This result can be combined with the relation~\eqref{eq:D_y} 
to yield the expression for the electric field inside the FLC layer
\begin{equation}
  \label{eq:E_y}
  E=\frac{\varepsilon_0 E_0 -\eta P_y+
    \avr{P_y\varepsilon_{yy}^{-1}}-P_{y}\avr{\varepsilon_{yy}^{-1}}}{%
\varepsilon_0\varepsilon_{yy}
[\avr{\varepsilon_{yy}^{-1}}+\eta]
}.
\end{equation}

For a uniform director distribution,
Eq.~\eqref{eq:E_y}
can be simplified giving the relation
\begin{equation}
  \label{eq:E_y-uni}
  E=\frac{E_0 -\eta P_y/\varepsilon_0}{%
1+\varepsilon_{yy}\eta
}
\end{equation}
recently obtained in Ref.~\cite{Blinov:pre:2005}
for the case of identical substrates.
Evidently, this relation implies that
the electric field inside the FLC layer, $E$,
deviates from the applied electric field, $E_{0}$,
due to the presence of the aligning substrates.
As it was emphasized in Refs.~\cite{Blinov:pre:2002,Callaghan:pre:2003},
this effect plays an important part in
describing the V-shaped switching mode.

Alternatively, Eq.~\eqref{eq:E_y-uni} can be used
to introduce the effective electrostatic
free energy
\begin{equation}
  \label{eq:f_eff}
  2 f_{\eff}= -(1+\eta\varepsilon_{yy})^{-1}
[
\varepsilon_0\varepsilon_{yy} E_0^2+2 E_0 P_y -\eta P_y^2/\varepsilon_0
]
\end{equation}
which is defined so as to meet the condition 
\begin{equation}
  \label{eq:eff-rel}
-\pdr{f_{E}(\phi,E)}{\phi}=
\pdr{\varepsilon_{yy}}{\phi} \varepsilon_0 E^2/2+
\pdr{P_{y}}{\phi} E
= -\pdr{f_{\eff}(\phi,E_0)}{\phi} 
\end{equation}
expressing equivalence of the torques
computed from the free energy~\eqref{eq:f_E}
and the effective energy~\eqref{eq:f_eff}.
The last term in the square brackets on the right hand side of
Eq.~\eqref{eq:f_eff} represents
the energy of self-interacting polarization charges.

Now we 
neglect the dielectric anisotropy of FLC,
$\varepsilon_{yy}\approx \varepsilon_{\FLC}=\varepsilon_{\perp}$,
and consider the energy
\begin{equation}
  \label{eq:f-total}
F\equiv f d = f_{\eff} d + W_{\anc}  
\end{equation}
defined as a sum of the effective and the anchoring energies.
After substituting Eqs.~\eqref{eq:anchoring-uni} and~\eqref{eq:f_eff}
into the energy~\eqref{eq:f-total},
it can be seen that there are two
effects caused by the aligning layers:
(a)~reduction of the applied electric field:
$E_{0}\to [1+\eta\varepsilon_{\FLC}]^{-1} E_{0}$,
and
(b)~renormalization
of the nonpolar anchoring strength induced by
the polarization self-interaction:
$W\to W-\eta   [1+\eta\varepsilon_{\FLC}]^{-1} \varepsilon_0^{-1}
P_{\spn}^2$. The latter is the effect that may result in breaking the
metastability condition~\eqref{eq:meta-stability}
thus leading to the formation of monostable structures
exhibiting the thresholdless
switching~\cite{Copic:pre:2002,Copic:pre:2003}.
Since such a possibility as well as V-shaped switching
is beyond the scope of this paper, 
we shall use the energy~\eqref{eq:f-total}
without changing notations for
the renormalized values of $E_{0}$ and $W$. 

The simple model of switching dynamics
in FLC cells can be formulated in terms of the
energy~\eqref{eq:f-total} which enters 
the dynamical equation for the azimuthal angle
\begin{equation}
  \label{eq:dyn-uni}
  \gamma\,\pdr{\phi}{t} = - \pdr{f}{\phi},
\end{equation}
where 
$\gamma$ is the rotational viscosity 
for reorientation on the smectic cone.
The model is based on the assumption of spatially uniform director field
that implies
neglecting elasticity effects and the coupling between the fluid flow
and the director.

Such uniform theory suggests applying
the viscosity-limited dynamics to
the switching process in SSFLC cells.
The simplest case where the anchoring energy is disregarded 
was originally considered in~\cite{Clark:apl:1980,Hands:apl:1982}
(see e.g.~\cite{Clark1:incoll:1991} for a review).
An attempt to fit experimental data using this simplified
model was recently made in~\cite{Essid:lc:2005}.

Subsequently, various modifications
of the model such as including 
an effective elastic-like term~\cite{Dahl:pra:1987}
and the anchoring energy~\cite{Pozhid:mm:1993,Sako:apl:1997,Callaghan:pre:2003}
have been employed to interpret experimental results.
The approach to grey levels in FLC displays based on the uniform
theory  is presented in~\cite{Pauwels:lc:2001}. 

In our case there are two characteristic time scales
\begin{equation}
  \label{eq:time-scales}
  t_{W}=\frac{\gamma d}{W},\quad
t_{E}=\frac{\gamma d}{ P_{\spn} V_{m}}
\end{equation}
and the governing equation~\eqref{eq:dyn-uni}
can be written in the following explicit form
\begin{align}
&
  \label{eq:dyn-phi}
  \pdr{\phi}{\tau} = - (r(\tau)+\cos\phi+w_p)\sin\phi,
\\
&
\label{eq:tau}
\tau=t/t_{W},\quad
r(\tau)=r_{e} v(\tau),
\end{align}
where  $r_{e}=t_{W}/t_{E}=P_{\spn} V_{m}/W$ 
is the \textit{driving voltage parameter}. 
Note that rescaling of time renormalizes the period of the waveform function
$v(\tau)$: $T\to T_{w}=T/t_{W}$.

Eq.~\eqref{eq:dyn-phi} presents the simplest model that can be used
to study the effects of asymmetry induced by the polar anchoring
energy. The characteristics of the hysteresis loop
depicted in Fig.~\ref{fig:hysteresis} will be of our
primary concern. 

Referring to Fig.~\ref{fig:hysteresis}, 
location of the peaks of the polarization reversal current 
is determined by two  switching voltage thresholds, 
$V_{+}= V_m v_{+}$
and $V_{-}=-V_{m} v_{-}$,  
where $v_{\pm}$ is the dimensionless \textit{switching voltage parameter},
characterizing the hysteresis loop for the normalized polarization
(\textit{polarization parameter}), $p=P_{y}/P_{\spn}=\cos\phi$.
Equivalently, width and shift of the loop
can be conveniently described by  
the dimensionless (normalized) voltage parameters
\begin{equation}
  \label{eq:coerc-shift}
  v_{c}=v_{+}+v_{-},
\quad
v_{\shf}=v_{+}-v_{-},
\end{equation}
where $v_{c}$ is
the \textit{voltage coercitivity} 
and $v_{\shf}$ is the \textit{voltage shift}. 

If $w_p=0$, the dynamical equation~\eqref{eq:dyn-phi}
is invariant under the symmetry transformation:
$r\to -r$ ($\tau\to\tau + T_{w}/2$) and $\phi\to\pi - \phi$ ($p\to
-p$). Under these circumstances, the hysteresis loop is symmetric and the voltage
shift vanishes, $v_{\shf}=0$. 
This is no longer the case when $w_p\ne 0$ and
the symmetry is broken.
The curves representing
temporal evolution of the normalized polarization, $p$, and its derivative with respect to
time driven by the sine-wave voltage at $w_p=0.2$ are shown in Fig.~\ref{fig:time-diagram}.

%%%%%%%%%%%%%%%%%%%%%%%%%%%%%%%%%%%%%%
\subsection{Switching voltages vs frequency: simplified model}
\label{subsec:no-achoring}
%%%%%%%%%%%%%%%%%%%%%%%%%%%%%%%%%%%%%%

It is instructive to examine first 
predictions of the extremely simple model
in which the terms describing the anchoring energy~\eqref{eq:anchoring-uni}
are disregarded and dynamics of the azimuthal angle
$\phi$ is governed by the simplified equation of motion 
\begin{equation}
  \label{eq:vos-phi}
\gamma\,\pdr{\phi}{t}=- P_{\spn}\, E_{0} \,v(t) \sin\phi.  
\end{equation}
Our task is to study the switching voltages
as functions of the driving voltage frequency, $f=1/T$.
We shall perform analysis of the model~\eqref{eq:vos-phi}
in the form suitable for subsequent generalization 
and clarify the assumptions underlying 
the results previously reported
in Refs.~\cite{Reyn:jpd:1989,Reyn:ferro:1991}.

We begin with 
introducing an auxiliary angular variable $u$
linked to the azimuthal angle
through the relations
\begin{equation}
  \label{eq:vos-u-var}
2u=\ln\left(
\frac{1+\cos\phi}{1-\cos\phi}
\right),
\quad
\cos\phi=\tanh(u) 
\end{equation}
and governed by the equation taken in the dimensionless form
\begin{align}
  \label{eq:vos-u-eq}
  \pdr{u}{\tau}=v(\tau),
\quad
\tau=t/t_{E}.
\end{align}
Solving Eq.~\eqref{eq:vos-u-eq} gives
the normal component of the polarization vector,
$P_{y}$, evolving in time as follows
\begin{align}
  \label{eq:vos-P-n}
  &
P_{y}(\tau)=P_{\spn}\cos\phi(\tau)=P_{\spn}\tanh(u(\tau)),
\\
&
\label{eq:vos-u-t}
u(\tau)=-u_0+w(\tau)=-u_0+\int_0^{\tau} v(\tau')\dd\tau',
\end{align}
where the initial condition $\cos\phi(0)=-\tanh u_0$
with $u_0>0$ means that the cell is initially in the down state. 

Periodicity of the polarization~\eqref{eq:vos-P-n} in time
requires the condition $w(T_e)=0$, 
where $T_e=T/t_{E}$ is the renormalized period,
to be fulfilled.
For bipolar switching with the waveform functions~\eqref{eq:sin-wv}-~\eqref{eq:squar-wv}, 
we have
\begin{align}
&
  \label{eq:vos-w-sin-wv}
  w_{\sine}(\tau)=[1-\cos(\omega_e\tau)]/\omega_e,
\quad
\omega_{e} = 2\pi/T_{e},
\quad
T_{e}=T/t_{E},
\\
&
\label{eq:vos-w-triang-wv}
w_{\triang}(\tau)= T_{e}
\begin{cases}
2 (\tau/T_{e})^2, & 0\le \tau/T_{e}\le 1/4,\\
-2 (\tau/T_{e})^2+2\tau/T_{e}-1/4, & 1/4\le \tau/T_{e}\le 3/4,\\
2 (\tau/T_{e}-1)^2, & 3/4\le \tau/T_{e}\le 1,
\end{cases}
\\
&
\label{eq:vos-w-squar-wv}
w_{\squar}(\tau)=T_{e}
\begin{cases}
\tau/T_{e}, & 0\le \tau/T_{e}< 1/2,\\
1-\tau/T_{e}, & 1/2\le \tau/T_{e}< 1,
\end{cases}
\end{align}
so that the periodicity condition is satisfied.

From Eq.~\eqref{eq:vos-u-t} the angular variable $u$ monotonically increases from
$-u_0$ to $u_{+}=-u_0+w(T_{e}/2)$ over the first half-period.
Complete switching occurs when $\cos\phi_{0}=-\tanh u_{0}\approx -1$ and
$\cos\phi_{+}=\tanh u_{+}\approx 1$.
In this case, the polarization
increases (decreases) passing zero at the instant of time $\tau_{+}$ ($\tau_{-}$). 
So, the switching voltage parameters, $v_{+}$ and $v_{-}$, 
can be defined as follows
\begin{align}
  \label{eq:vos-v-pm}
  v_{\pm}=\pm v(\tau_{\pm}),
\quad
u(\tau_{\pm})=0.
\end{align}
For sine-wave and triangular voltages,
Eq.~\eqref{eq:vos-v-pm} can be combined with
Eqs.~\eqref{eq:vos-u-t}-\eqref{eq:vos-w-triang-wv}
to yield the expressions for the switching voltages:
\begin{align}
&
  \label{eq:vos-vpm-sin}
  v^{(\sine)}_{\pm}=\sqrt{u_{0}\omega_{e}(2-u_{0} \omega_{e})},
\quad
u_{0} \omega_{e}\le 1,
\\
&
\label{eq:vos-vpm-triang}
v^{(\triang)}_{\pm}=
\begin{cases}
\sqrt{8u_{0}/T_{e}},& 0\le u_{0}/T_{e}\le 1/8,\\
\sqrt{2-8u_{0}/T_{e}},& 1/8\le u_{0}/T_{e}\le 1/4.
\end{cases}
\end{align}
 It would appear natural that 
the polarization in the up and down states 
is of the same magnitude but differ in sign.
So,
the parameter $u_{0}$ entering
the expressions~\eqref{eq:vos-vpm-sin} and~\eqref{eq:vos-vpm-triang} 
can be fixed by imposing the condition
$\cos\phi_{+}=-\cos\phi_{0}$ ($u_{+}=u_{0}$).
In this case, we have
\begin{align}
  \label{eq:vos-symm}
  2 u_0=w(T_{e}/2).
\end{align}
Since $w_{\sine}(T_e/2)=2/\omega_e$ and
$w_{\triang}(T_e/2)=T_e/4$,
we are led to the conclusion that
the switching voltages are frequency independent
and $v^{(\sine)}_{\pm}=v^{(\triang)}_{\pm}=1$.

As a way around this difficulty,
we can apply a cutoff procedure where the equilibrium
states are regarded as ``saturated'' states
characterized by the cutoff parameter
$u_s$, $\tanh u_{s}\approx 1$. 
The result is that
the relation
between the angular and polarization parameters
takes the modified form
\begin{align}
  \label{eq:vos-u-s}
  \cos\phi=
\begin{cases}
\tanh u,& |u|\le u_{s},\\
\sign(u)\tanh u_s\equiv \pm\cos\phi_{s},& |u|\ge u_{s}
\end{cases}
\end{align}
and the parameter $u_{0}$ is now given by
\begin{align}
  \label{eq:vos-u-0}
  2 u_0=
\begin{cases}
w(T_{e}/2),& T_{e}\le T_{s},\\
2 u_{s}=w(T_{s}/2),& T_{e}\ge T_{s}. 
\end{cases}
\end{align}
From Eq.~\eqref{eq:vos-u-0} it can be seen that
the cutoff parameter $u_s$ determines the boundary frequency,
$f_s=1/T_s$, separating the regions of low and high frequencies. 
In the low frequency regime with $f_e=1/T_e<f_s$,
the switching voltage parameters are given by
\begin{align}
  \label{eq:vos-vpm-fin}
  v^{(\sine)}_{\pm}=\sqrt{(2-\omega_{e}/\omega_s)\omega_{e}/\omega_s},
\quad
v^{(\triang)}_{\pm}=\sqrt{f_{e}/f_{s}},
\end{align}
where $\omega_s=2\pi f_s$.

Note that, similar to the switching voltages, 
the switching time parameters, $\tau_{+}/T_e$ and $\tau_{-}/T_e$,
are frequency independent, $\tau_{+}/T_e=1-\tau_{-}/T_e=1/4$,
at $f_e>f_s$ for all three wave forms~\eqref{eq:sin-wv}-\eqref{eq:squar-wv}. 
For the square-wave driving voltage of low frequency,
these parameters assume linear dependence on the frequency: 
$\tau_{+}/T_e=1-\tau_{-}/T_e=f_e/(4 f_s)$.

%%%%%%%%%%%%%%%%%%%%%%%%%%%%%
\section{Switching dynamics}
\label{sec:switch-dyn}
%%%%%%%%%%%%%%%%%%%%%%%%%%%%%

From analysis presented in Sec.~\ref{subsec:no-achoring} 
it can be inferred
that properties of switching dynamics
are predominately determined by a certain class of  
periodic solutions (cycles) of the governing equation.
Specifically, we have used the symmetry condition
$p_{+}=-p_{-}$, where $p_{+}$ ($p_{-}$) is the polarization parameter
of the up (down) state, in combination with the cutoff procedure. 

Now we turn back to the model~\eqref{eq:dyn-phi}
and study how the anchoring energy influences the dynamics of
switching. As in the preceding section, our attention will be focused on
the frequency dependence of the switching voltages in the regime
of complete switching.

By contrast to Eq.~\eqref{eq:vos-phi}, the dynamical
equation~\eqref{eq:dyn-phi} is generally not exactly solvable.
Additional difficulties emerge 
in analysis of the solutions describing the switching process
that need to be periodic in time and will be referred to as the
\textit{cycles}. 
As we shall be seeing the very existence of cycles does not follow
from Eq.~\eqref{eq:dyn-phi} immediately and the periodicity
conditions may require somehow involved considerations. 

In this section we present an analytical approach
to study the cycles based on approximating
the dynamical equation for the angular variable $u$
defined in Eq.~\eqref{eq:vos-u-var}.
By applying this method, 
the case of square wave voltages can be treated analytically,
whereas some relatively simple numerical analysis is needed
for other wave forms.  

As a first step, we shall write the equation for
the angular variable $u$
\begin{equation}
  \label{eq:dyn-u}
  \pdr{u}{\tau}\equiv\dot{u} = \tanh(u)+ [r(\tau)+w_p],
\quad
p\equiv \cos\phi=\tanh(u)
\end{equation}
deduced from Eq.~\eqref{eq:dyn-phi} by 
using Eq.~\eqref{eq:vos-u-var} to make
the change of variables: $\phi\to u$.
Then we apply the piecewise linear approximation for hyperbolic tangent
\begin{equation}
  \label{eq:tanh-approx}
  \tanh(u)\to L(u)=
\begin{cases}
u, & |u|<1,\\
\sign(u),& |u|\ge 1
\end{cases}
\end{equation}
and obtain the approximate dynamical equation
\begin{equation}
  \label{eq:dyn-u-approx}
  \dot{u} = L(u)+ [r(\tau)+w_p]
\end{equation}
that will be the starting point for our subsequent considerations.
Eq.~\eqref{eq:dyn-u-approx} 
retains the symmetry of Eq.~\eqref{eq:dyn-u}
discussed at the end of Sec.~\ref{subsec:energy}.
It can be readily checked 
that the transformation
\begin{equation}
  \label{eq:symmetry}
  \tau\to\tau-T_{w}/2,\quad
u\to - u,\quad
w_{p}\to -w_{p}
\end{equation}
keeps both of these equations intact.

It turned out that, as far as the cycles are concerned, 
the numerical results computed from Eq.~\eqref{eq:dyn-u}
are essentially the same as compared with 
the predictions of Eq.~\eqref{eq:dyn-u-approx}.
In our calculations the relative error 
was found to be well below $0.1$ per cent.

Accuracy of the approximation is demonstrated in
Fig.~\ref{fig:time-diagram}. 
It is seen that the difference between the solid and dashed lines
becomes noticeable only with a rise in numerical integration error
as the interval of integration increases.

As in Sec.~\ref{subsec:no-achoring}, suppose that the cell is
initially in the down state with $u(0)=-u_{0}$ and $u_{0}>1$.
After the first half-period time, the external voltage drives the cell
into the up state with $u(T_{w}/2)=u_{+}$. This process might be
called \textit{switching up} and we consider it complete if $u_{+}>1$.
By analogy, complete \textit{switching down} takes place over the second half-period
time when the cell goes from the up state with $u(T_{w}/2)=u_{+}$ to the
down state with $u(T_{w})=-u_{-}$ and $u_{-}>1$.

Mathematically, this can be described in terms
of the half-period mappings
\begin{align}
&
  \label{eq:mapping-gen}
  \Psi_{+}: u_{0}=-u(0)\Mapsto u_{+}=u(T_{w}/2),
\nonumber
\\
&
  \Psi_{-}: u_{+}=u(T_{w}/2)\Mapsto u_{-}=-u(T_{w}),
\end{align}
where $\Psi_{+}$ and $\Psi_{-}$ are determined by
temporal evolution of the angular variable $u$ over
the first and second half-period, respectively.

Owing to the symmetry~\eqref{eq:symmetry},
$\Psi_{-}$ can be obtained from $\Psi_{+}$
by changing the sing of the polar anchoring parameter
$w_p$. So, in the subsequent section, we concentrate on 
the half-period mapping $\Psi_{+}$. 

%%%%%%%%%%%%%%%%%%%%%%%%%%%
\subsection{Half-period mappings and cycles}
\label{subsec:half-period-mapping}
%%%%%%%%%%%%%%%%%%%%%%%%%%%

We restrict ourselves to the first half-period
and study in detail how the angular parameter $u$ evolves
in time. 
For complete switching up, the solution
can be written in the following general form
\begin{align}
&
  \label{eq:u-sol}
  u(\tau)=
\begin{cases}
F_{0}(\tau), & \tau<\tau_{0},\\
F(\tau),& \tau_{0}\le\tau\le \tau_{1},\\
F_{+}(\tau),& \tau_{1}<\tau\le T_{w}/2,
\end{cases}
\end{align}
where $0<\tau_0<\tau_1\le T_{w}/2$ are defined by the conditions
\begin{equation}
  \label{eq:tau-01}
u(\tau_0)=-1,\quad
u(\tau_1)=1.
\end{equation}
It is not difficult to solve Eq.~\eqref{eq:dyn-u-approx}
and derive
the explicit formulas for the functions that enter 
the solution~\eqref{eq:u-sol}.
The result is
\begin{align}
&
\label{eq:F0}
F_{0}(\tau)=-u_0+(w_p-1)\tau+R_{0}(\tau),
\\
&
\label{eq:F}
F(\tau)=\exp(\tau-\tau_{0})(w_p-1)-w_p+
\exp(\tau)(R(\tau_{0})-R(\tau)),
\\
&
\label{eq:Fplus}
F_{+}(\tau)=1+(w_p+1)(\tau-\tau_{1})+R_{0}(\tau)-R_{0}(\tau_{1}),   
\end{align}
where
\begin{align}
&
  \label{eq:R0}
  R_{0}(\tau)=\int_{0}^{\tau} r(\tau')\dd\tau',
%\\
%&
%\label{eq:R}
\quad
  R(\tau)=-\int_{0}^{\tau} \exp(-\tau')\,r(\tau')\dd\tau'.
\end{align}

From Eqs.~\eqref{eq:u-sol}-\eqref{eq:Fplus}
the half-period mapping $\Psi_{+}$
from $u_{0}$ to $u_{+}$ can be derived in
the parameterized form
\begin{subequations}
  \label{eq:Psi-plus-gen}
\begin{align}
\Psi_{+}:\quad
&
  \label{eq:G0}
  u_{0}-1=G_{0}(\tau_{0})=(w_p-1)\tau_{0}+R_{0}(\tau_{0}),
\\
&
\label{eq:Gplus}
u_{+}-1=G_{+}(\tau_{1})=(w_p+1)(T_{w}/2-\tau_{1})+R_{0}(T_{w}/2)-R_{0}(\tau_{1}),
\\
&
\label{eq:g-coupl}
g_{-}(\tau_{0})=g_{+}(\tau_1),
\quad
g_{\pm}(\tau)=\exp(-\tau)(w_p\pm 1) + R(\tau),   
\end{align}
\end{subequations}
where the parameters $\tau_0$ and $\tau_1$ are defined by the two conditions~\eqref{eq:tau-01}.
The first condition $u(\tau_0)=F_{0}(\tau_{0})=-1$ and the relation
$u(T_{w}/2)=F_{+}(T_{w}/2)=u_{+}$ give $u_{0}$ and $u_{+}$ as
a function of $\tau_{0}$ and $\tau_{1}$, respectively (see
Eqs.~\eqref{eq:G0} and~\eqref{eq:Gplus} for the corresponding expressions).
The parameters $\tau_0$ and $\tau_1$ are coupled through
the equation~\eqref{eq:g-coupl} 
which is an immediate consequence of the second condition
from Eq.~\eqref{eq:tau-01}: $u(\tau_1)=F(\tau_{1})=1$.

Changing the sign of $w_p$
in Eqs.~\eqref{eq:G0}-\eqref{eq:g-coupl}
gives the formulas for the half-period mapping $\Psi_{-}$.
Composition of the two mappings 
\begin{align}
  \label{eq:Psi-minus}
\Psi=\Psi_{-}\circ\Psi_{+},
\quad
  \Psi_{-}=\Psi_{+}\Bigl\lvert_{w_p\to\, -w_p}
\end{align}
relates the values of $u$ at the end points of the period: 
$u(0)=-u_0\to u(T_{w})=-u_{-}$.

From the periodicity condition $u_0=u_{-}$ it follows
that, for cycles, $u_0$ is the fixed point of $\Psi$.
So, we have the fixed point equations
\begin{subequations}
  \label{eq:fixed-points-gen}
\begin{align}
&
  \label{eq:fixed-minus}
  \Psi(q_{-}^{(\st)})=q_{-}^{(\st)},
\quad
q_{-}^{(\st)}\equiv u_{-}^{(\st)}-1=u_{0}^{(\st)}-1,
\\
&
\label{eq:fixed-plus}
\Psi_{+}(q_{-}^{(\st)})=q_{+}^{(\st)}\equiv u_{+}^{(\st)}-1
\end{align}
\end{subequations}
characterizing the up and down states of the cycle.
The fixed point then can be found by solving Eq.~\eqref{eq:fixed-minus}.
It provides the polarization parameter for the down state
of the cycle: $p_{-}^{(\st)}=-\tanh u_{-}^{(\st)}$, whereas
Eq.~\eqref{eq:fixed-plus} gives $u_{+}^{(\st)}$ and
the polarization parameter of the up state:
$p_{+}^{(\st)}=\tanh u_{+}^{(\st)}$.

Given the cycle Eqs.~\eqref{eq:dyn-u}-\eqref{eq:R0}
with $u_0=u_{-}^{(\st)}$ yield $u(\tau)$ for 
switching up during the first half-period,
$0\le\tau\le T_{w}/2$. The reversal polarization current
reaches a maximum at the instant of time
$\tau_{+}$ where $\ddot{p}(\tau_{+})=0$.

For the second half-period time,
$T_{w}/2\le\tau\le T_{w}$,
Eqs.~\eqref{eq:dyn-u}-\eqref{eq:R0}
with $u_0=u_{+}^{(\st)}$ and $w_p\to -w_p$
give $-u(\tau-T_{w}/2)$. The current
is now peaked at $\tau=T_{w}/2+\tau_{-}$. 
From the condition $\ddot{p}=0$
and Eq.~\eqref{eq:dyn-u} we can derive the equations
for $\tau_{+}$ and $\tau_{-}$
\begin{align}
&
  \label{eq:tau-pm}
  \dot{u}(\tau_{\pm})
\bigl[
2\dot{u}(\tau_{\pm})\tanh u(\tau_{\pm})-1
\bigr]\approx
\dot{u}(\tau_{\pm})
\bigl[
2\dot{u}(\tau_{\pm})u(\tau_{\pm})-1
\bigr]
=
\dot{r}(\tau_{\pm}),
\\
&
\label{eq:dot-u}
\dot{u}(\tau_{\pm})=u(\tau_{\pm})+[r(\tau_{\pm})\pm w_p],
\end{align}
where 
$0<\tau_{\pm}<T_{w}/2$,
$u(\tau_{+})=F(\tau_{+})$ and
$u(\tau_{-})=F(\tau_{-})\lvert_{w_p\to\, -w_p}$.
Then the switching voltages are given by
\begin{align}
  \label{eq:r-pm}
  r_{\pm}=\pm r(\tau_{\pm})=\pm r_e v_{\pm},
\end{align}
where $v_{\pm}=v(\tau_{\pm})$ is the switching voltage parameter.

\begin{figure*}[!tbh]
%\vskip5mm
\centering
\resizebox{150mm}{!}{\includegraphics*{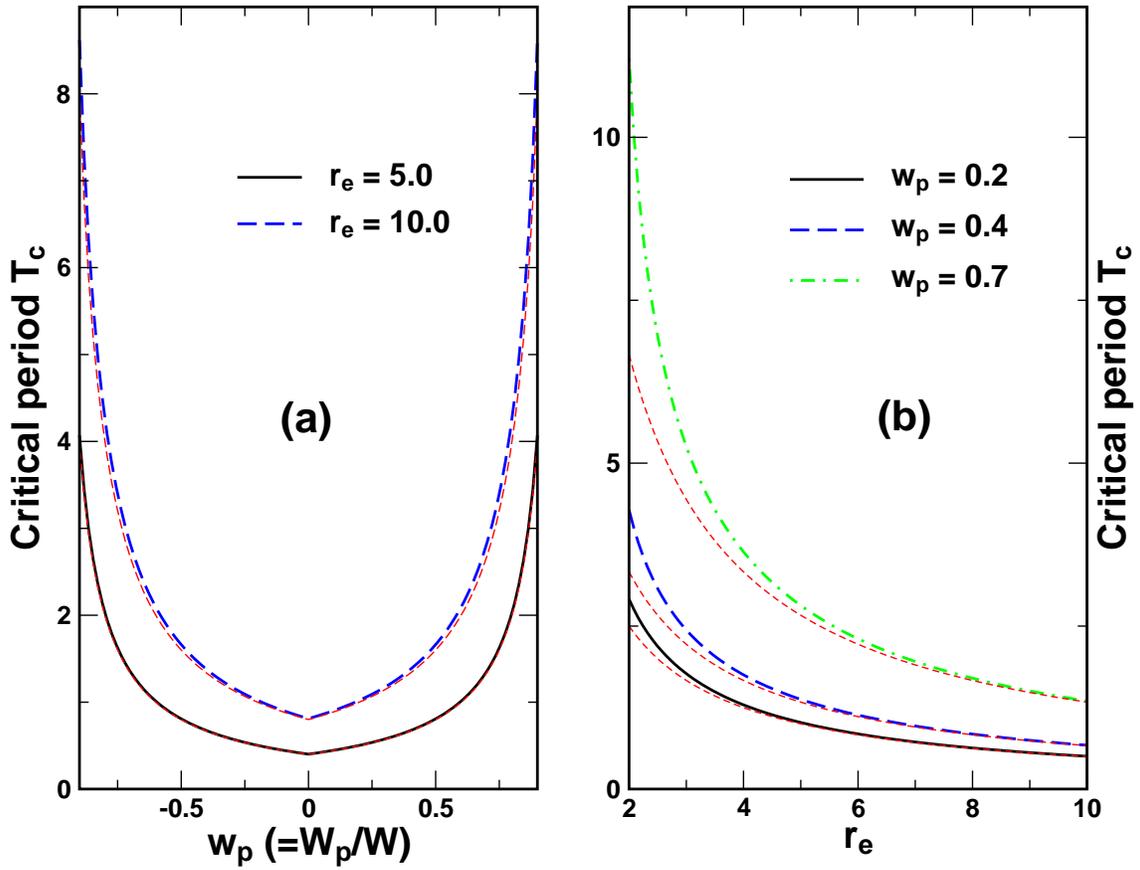}}
\caption{%
Critical period for square wave voltage as a function
of (a)~polar anchoring parameter, $w_p=W_p/W$ 
and (b)~voltage amplitude parameter, $r_e=P_{\spn} V_{\mathrm m}/W$.
Thin dashed lines are computed from the approximate
formula~\eqref{eq:T-c-approx-sqr}.
}
\label{fig:Tc-sqr}
\end{figure*}

%%%%%%%%%%%%%%%%%%%%%%%%%%
\subsection{Square wave voltage}
\label{subsec:square-wave-voltage}
%%%%%%%%%%%%%%%%%%%%%%%%%%

Analytical results of the preceding section
depend on the wave form through the functions
defined in Eq.~\eqref{eq:R0}.
The case of square wave voltages
can be treated by replacing $w_p$ with $w_{p}+r_e$
and setting $R_{0}$ and $R$ equal to zero.
The result for the functions that enter the half-period mappings
(see Eqs.~\eqref{eq:G0} and~\eqref{eq:Gplus}) 
and the coupling equation~\eqref{eq:g-coupl} is
\begin{subequations}
  \label{eq:Psi-param-sqr}
\begin{align}
\Psi_{\pm}:\quad
&
  \label{eq:G0-sqr}
  G_{0}=(\pm w_p+r_e-1)\tau_{0},
\\
&
\label{eq:Gplus-sqr}
G_{+}=(\pm w_p+r_e+1)(T_{w}/2-\tau_{1}),
\\
&
\label{eq:tau01-sqr}
\tau_{1}=\tau_{0}+\gamma_{\pm}=\tau_{0}+\ln\beta_{\pm},
\quad
\beta_{\pm}=\frac{r_e\pm w_p+1}{r_e\pm w_p-1}.   
\end{align}
\end{subequations}
By using Eqs.~\eqref{eq:G0-sqr}-\eqref{eq:tau01-sqr}
along with the formulas~\eqref{eq:G0}-~\eqref{eq:Gplus}
the parameters $\tau_0$ and $\tau_1$ 
can be eliminated to 
yield
the half-period mappings, $\Psi_{+}$ and $\Psi_{-}$, 
as the linear functions given by
\begin{subequations}
  \label{eq:Psi-explicit-sqr}
\begin{align}
&
  \label{eq:Psi-plus-sqr}
  q_{+}=\Psi_{+}(q_{0})=\alpha_{+}-\beta_{+} q_0\equiv q_1^{(+)} (1-q_0/q_0^{(+)}),
\\
&
  \label{eq:Psi-minus-sqr}
  q_{-}=\Psi_{-}(q_{+})=\alpha_{-}-\beta_{-} q_{+}\equiv q_1^{(-)} (1-q_+/q_0^{(-)}),
\end{align}
\end{subequations}
where $q_i\equiv u_i-1$ and
\begin{equation}
  \label{eq:alpha-sqr}
q_1^{(\pm)}\equiv \alpha_{\pm}=(\pm
w_p+r_e+1)(T_{w}/2-\gamma_{\pm}),
\quad
q_0^{(\pm)}=\alpha_{\pm}/\beta_{\pm}.  
\end{equation}

Now that the expressions for the half-period mappings derived
we discuss some of the most important consequences.
The first point to address concerns the regime of complete switching.

This regime occurs provided the parameters
$q_{0}=u_0-1$, $q_{+}=u_{+}-1$ and $q_{-}=u_{-}-1$
are all positive. Since $q_{0}=G_{0}(\tau_0)$ with $\tau_{0}>0$,
from Eq.~\eqref{eq:G0-sqr} we obtain the inequality
\begin{equation}
  \label{eq:switch-r-sqr}
  r_e>1+|w_p|
\end{equation}
as the complete switching condition for the driving voltage parameter,
$r_e=P_{\spn} V_{\mathrm m}/W$.

It is evident from Eqs.~\eqref{eq:Psi-plus-sqr}
and~\eqref{eq:Psi-minus-sqr} that the parameters $q_{+}$ and $q_{-}$
can be positive only if $q_1^{(\pm)}>0$.
The latter and Eq.~\eqref{eq:alpha-sqr} give the condition
of complete switching for the period of driving voltage
\begin{equation}
  \label{eq:T-min-sqr}
T_{w}>T_{\mn}=2\max(\gamma_{+},\gamma_{-})=
2\ln\frac{r_e-|w_p|+1}{r_e-|w_p|-1}
\end{equation}
restricting the driving frequency $1/T_{w}$ from above.

The conditions of complete switching~\eqref{eq:switch-r-sqr}
and~\eqref{eq:T-min-sqr} define the 
complete switching thresholds for
the amplitude and the period of the applied voltage.
According to estimates that will discussed in Sec.~\ref{sec:experim},
the voltage parameter $r_e$ is typically much larger than unity
and the asymptotic expression 
\begin{equation}
  \label{eq:T-min-approx-sqr}
  T_{\mn}\approx
4 (r_e^{-1}+|w_p| r_e^{-2})
\end{equation}
can be used as a good approximation for 
the threshold period~\eqref{eq:T-min-sqr}
at $r_e>10$.

In general, the complete switching conditions
are insufficient for cycles to exist.
In other words, we need to impose more stringent constraints
in order that the composition of mappings~\eqref{eq:Psi-minus}
has fixed points.

In our case, the mapping $\Psi=\Psi_{-}\circ\Psi_{+}$: $q_0\to q_{-}$ is linear and
the parameter $q_{0}$ varies in the range from zero to $q_0^{(+)}$.
The fixed point exists only if the function $q-\Psi(q)$ takes the
values of opposite sign at the endpoints of the interval $[0, q_0^{(+)}]$: 
$q=0$ and $q=q_0^{(+)}$.
Since 
 \begin{equation}
  \label{eq:Psi-bound-sqr}
  \Psi(0)=
\bigl[
q_0^{(-)}-q_1^{(+)}
\bigr]
q_1^{(-)}/q_0^{(-)},
\quad
  q_0^{(+)}-\Psi(q_0^{(+)})=q_0^{(+)}-q_1^{(-)},
\end{equation}
the cycle condition $\Psi(0) [q_0^{(+)}-\Psi(q_0^{(+)})] >0$
can be conveniently written as the inequality 
\begin{equation}
  \label{eq:cycle-sqr}
  P_{+}P_{-}>0,
\quad
P_{\pm}=q_0^{(\pm)}-q_1^{(\mp)},
\end{equation}
where the parameters $q_{0}^{(\pm)}$ and $q_1^{(\pm)}$
are defined in Eq.~\eqref{eq:alpha-sqr}
and linearly depend on $T_{w}$. 

It can be shown that $P_{+}$ and $P_{-}$ are both negative
in the region where the cycles come into play.
For this region of cycles, we have
\begin{equation}
  \label{eq:unstable-fixed}
  P_{\pm}<0\quad\text{at}\quad
T_{w}>T_{c},
\end{equation}
where $T_c$ is the critical period given by
\begin{equation}
  \label{eq:T-c-sqr}
  T_c=\ln\frac{(r_e+1)^2-w_p^2}{(r_e-1)^2-w_p^2}+
r_e\,(1-|w_p|)^{-1}\,\ln\frac{r_e^2-(1-|w_p|)^2}{r_e^2-(1+|w_p|)^2}.
\end{equation}
Thus there are no cycles in the high frequency region
above the critical frequency $1/T_c$. 

Similar to $T_{\mn}$, 
the critical period $T_c$ can be conveniently approximated
by the simplified asymptotic formula
\begin{equation}
  \label{eq:T-c-approx-sqr}
  T_c\approx 4 (1-|w_p|)^{-1} r_e^{-1}
\end{equation}
in the region of high voltage parameters.
Clearly, $T_c$ declines as the voltage amplitude increases,
whereas the critical period becomes divergent when 
the magnitude of the polar anchoring
parameter approaches unity.
The curves depicted in Fig.~\ref{fig:Tc-sqr}
illustrate these effects and accuracy of the high voltage approximation.

An important point is that, under
the cycle condition~\eqref{eq:unstable-fixed},
the derivative of $\Psi$,
$\Psi'(q)=q_1^{(+)}q_1^{(-)} \bigl[q_0^{(+)}q_0^{(-)}\bigr]^{-1}$,
is greater than unity. 
Therefore, the fixed points and the cycles are unstable.
Note that stability of
the cycles requires $P_{+}$ and $P_{-}$ to be both positive.

For the half-period mappings of 
the form~\eqref{eq:Psi-plus-sqr} and~\eqref{eq:Psi-minus-sqr}, 
it is straightforward to deduce the expressions for
the parameters
$q_{+}^{(+)}$ and $q_{-}^{(\st)}$ 
\begin{equation}
  \label{eq:q-st-sqr}
  q_{\pm}^{(\st)}=
\frac{\alpha_{\pm}-\beta_{\pm}\alpha_{\mp}}{1-\beta_{+}\beta_{-}}
=\frac{q_1^{(\pm)}q_0^{(\mp)} P_{\pm}}{q_0^{(+)}q_0^{(-)}-q_1^{(+)}q_1^{(-)}}
\end{equation}
that characterize the up and down states of the cycles.
Solving Eq.~\eqref{eq:tau-pm} gives
the switching time parameters $\tau_{+}$ and $\tau_{+}$
\begin{align}
&
  \label{eq:tau-pm-sqr}
  \tau_{\pm}=q_{\mp}^{(\st)}/(r_e\pm w_p-1)+\delta\tau_{\pm},
\\
&
\label{eq:dlt-tau-sqr}
\delta\tau_{\pm}=\ln\frac{r_e\pm w_p+\sqrt{(r_e\pm w_p)^2+2}}{%
2(r_e\pm w_p-1)
}
\end{align}
that determine location of the reversal current peaks
in time during the first and second half-period
(switching up and down).
The formulas
\begin{align}
&
  \label{eq:u-st-approx-sqr}
  u_{\pm}^{(\st)}\approx
T_{w} (1\mp w_p)
\bigl[
r_e-(1\pm w_p)^2 r_e^{-1}
\bigr]/4
+2 (1\pm 2 w_p) r_e^{-2}/3,
\\
&
\label{eq:tau-pm-approx-sqr}
  \tau_{\pm}\approx
T_{w} (1\pm w_p)
\bigl[
1+(1\mp w_p) r_e^{-1}
\bigr]/4
\mp 4 w_p r_e^{-3} /3 
\end{align}
describe asymptotic behavior
of $u_{\pm}^{(\st)}$ and $\tau_{\pm}$
in the region of high voltage parameters
where $r_e\ge 10$.

As is expected from the symmetry~\eqref{eq:symmetry},
we have $u_{+}^{(\st)}=u_{-}^{(\st)}$
and $\tau_{+}=\tau_{-}$
in the limit of zero polar anchoring with $w_p=0$.
This is also the case where the difference
between the critical period $T_c$ and the threshold period
$T_{\mn}$ disappears.

Interestingly, at $w_p=0$, the leading term of the asymptotic
expansion~\eqref{eq:u-st-approx-sqr}
[the first term in the square brackets on the right hand side of
Eq.~\eqref{eq:u-st-approx-sqr}]
can be derived from the symmetry condition~\eqref{eq:vos-symm}
discussed in Sec.~\ref{subsec:no-achoring}.
More generally, Eqs.~\eqref{eq:u-st-approx-sqr}
and~\eqref{eq:tau-pm-approx-sqr} indicate
that,
quite similar to the high frequency regime 
considered in Sec.~\ref{subsec:no-achoring},
the parameters
$u_{\pm}^{(\st)}/T_{w}$ and $\tau_{\pm}/T_{w}$
are almost frequency independent
having only a weak linear dependence on 
the frequency $1/T_{w}$.

The effects of the polar anchoring induced asymmetry
depend on the polarity determined by the sign of $w_p$.
At the critical point with $T_{w}=T_{c}$
and $w_p>0$ [$w_p<0$],
we have $P_{+}(T_c)=0$ [$P_{-}(T_c)=0$] 
and the parameter $q_{+}^{(\st)}$ [$q_{-}^{(\st)}$]
vanishes.

In the case where $w_p>0$ and the down state
is metastable in the absence of applied voltage, for the cycles,
the polarization parameter of the down state $p_{-}^{(\st)}$
appears to be higher than that of the up state 
$p_{+}^{(\st)}$, $p_{-}^{(\st)}>p_{+}^{(\st)}$.
An important consequence of this is that the switching time 
$\tau_{+}$ is longer than $\tau_{-}$, 
$\tau_{+}>\tau_{-}$ and $(\tau_{+}-\tau_{-})/T_{w}\approx w_p/2$.
So, at $w_p>0$, 
the model predicts that,
in cycles driven by the square wave voltage,  
switching down is faster than switching up.
Note that the results for  reversed sign of $w_p$,
can be obtained by 
interchanging the up and down states. 
 
%%%%%%%%%%%%%%%%%%%%%%%%%%%%%%%
\subsection{Sine-wave voltage: numerical analysis}
\label{subsec:sine-wave}
%%%%%%%%%%%%%%%%%%%%%%%%%%%%%%%

Now we extend the 
analysis of the previous section to the cases
of sine-wave and triangular voltages.
More specifically, we shall present the results
for the case of the sine-wave voltage.
The triangular driving voltage can be treated in just the same way 
but this involves rather cumbersome expressions. 
So, our subsequent 
considerations are equally applicable to both of the waveforms.

As a first step,
we perform integrals in Eq.~\eqref{eq:R0} 
for the waveform function~\eqref{eq:sin-wv}
and deduce the expressions for the functions 
\begin{align}
&
  \label{eq:R-sine}
R(\xi)=r_e\exp(-\xi/\omega_{w})
\bigl[
\omega_{w} \cos\xi+
\sin\xi
\bigr]/(1+\omega_{w}^2),
\\
&
  \label{eq:R0-sine}
  R_{0}(\xi)=r_e 
\bigl[
1-\cos\xi
\bigr]/\omega_{w},\quad
\xi=\omega_{w}\tau, 
\end{align}
where $\omega_{w}=2\pi/T_{w}=\omega t_{W}$
is the renormalized frequency (frequency parameter). 
Substituting Eqs.~\eqref{eq:R-sine} and~\eqref{eq:R0-sine}
into Eqs.~\eqref{eq:G0}-\eqref{eq:g-coupl}
gives the half-period mapping $\Psi_{+}$
in the following parameterized form
\begin{subequations}
  \label{eq:Psi-plus-sine}
\begin{align}
\Psi_{+}:\quad
&
  \label{eq:q0-sine}
  q_0\equiv\omega_{w}(u_{0}-1)=(w_p-1)\xi_{0}+r_e
\bigl[
1-\cos\xi_0
\bigr]
\equiv G_0^{(\sine)}(\xi_0),
\\
&
\label{eq:qplus-sine}
q_{+}\equiv \omega_{w}(u_{+}-1)=(w_p+1)(\pi-\xi_{1})+r_e
\bigl[
1+\cos\xi_1
\bigr]
\equiv G_{+}^{(\sine)}(\xi_1),
\\
&
\label{eq:g-coupl-sine}
g_{-}(\xi_{0})=g_{+}(\xi_1),
\quad
g_{\pm}(\xi)=\exp(-\xi/\omega_{w})(w_p\pm 1) +R(\xi),   
\end{align}
\end{subequations}
where the parameters $\xi_0=\omega_{w}\tau_{0}$ and
$\xi_1=\omega_{w}\tau_{1}$ are ranged between zero and $\pi$.
[Note that the parameters $q_i$ differ from those defined in 
Sec.~\ref{subsec:square-wave-voltage}
by the factor $\omega_{w}$.]

As opposed to the case of square wave voltage, 
the parameters $\xi_0$ and $\xi_1$ cannot be eliminated.
We begin with locating the values of these parameters
that represent the regime of complete switching 
where $q_0$ and $q_{+}$ are non-negative, $q_{0,\,+}\ge 0$.

Our first remark concerns behavior of
the functions, $G_{0,\,+}$ and $g_{\pm}$, that enter the mapping $\Psi_{+}$.
The time derivatives of these functions 
\begin{align}
&
  \label{eq:der-G0-plus}
  \pdr{G_{0,\,+}(\tau)}{\tau}=-1\pm(w_p+r(\tau)),
\\
&
\label{eq:der-g-pm}
\pdr{g_{\pm}(\tau)}{\tau}=-\exp(-\tau)
\bigl(
w_p\pm 1+r(\tau)
\bigr)  
\end{align}
show that over the first half-period time $G_{+}$ and $g_{+}$
are monotonically decreasing functions of $\tau$.

By contrast, the functions $G_0$ and $-g_{-}$ exhibit nonmonotonic
behavior.  As $\tau$ increases, they first decay  reaching the
minimum located at $\tau_{\mn}$ and then grow up to the maximum at
$\tau=\tau_{\mx}$ decreasing at $\tau>\tau_{\mx}$.
The points $\tau_{\mn,\,\mx}$  are determined by the equation
$\dot{G}_{0}(\tau)=0$ [or, equivalently, $v(\tau)=(1-w_p)/r_e$]
and $G_{0}(\tau_{\mn})<0$.

For the sine-wave voltage, $\xi_{\mn}=\arcsin[(1-w_p)/r_e]$ and
$\xi_{\mx}=\pi-\xi_{\mn}$. The function $G_0^{(\sine)}$
monotonically increases from $G_0^{(\sine)}(\xi_{\mn})<0$
to $G_0^{(\sine)}(\xi_{\mx})$
and it passes through zero, $G_0^{(\sine)}(\xi_0^{(0)})=0$, 
before the waveform function reaches a maximum
only if $G_0^{(\sine)}(\pi/2)>0$.
The latter provides the complete switching condition for the driving
voltage parameter
\begin{equation}
  \label{eq:switch-r-sine}
  r_e>(1+|w_p|)\,\pi/2.
\end{equation}
Interestingly, replacing $r_e$ with $\pi r_e/2$
in Eq.~\eqref{eq:switch-r-sine}
recovers the result for the square wave voltage
given by Eq.~\eqref{eq:switch-r-sqr}.

The largest value of $\xi_0$,
$\xi_0=\xi_0^{(+)}$, can be found by solving 
the coupling equation~\eqref{eq:g-coupl-sine}
at $\xi_1=\pi$.
Owing to the inequality $\xi_0^{(+)}>\xi_0^{(0)}$,
the complete switching  condition
for the frequency is given by 
\begin{equation}
  \label{eq:switch-gen-sine}
  g_{-}(\xi_{0}^{(0)})>g_{+}(\pi).
\end{equation}
It can be checked that Eq.~\eqref{eq:T-min-sqr}
with $r_e$ replaced by $2 r_e /\pi$
provides a good estimate for $T_{\mn}$
defined by the condition~\eqref{eq:switch-gen-sine}.

At this stage we have found
that the endpoints of
the interval for $\xi_0$:
$\xi_0^{(0)} \le \xi_0\le \xi_0^{(+)}$,
can be evaluated as solutions of the equations
\begin{align}
  \label{eq:xi0plus0-sine}
  g_{-}(\xi_{0}^{(+)})=g_{+}(\pi),
\quad
G_0^{(\sine)}(\xi_0^{(0)})=0.
\end{align}
According to the coupling equation~\eqref{eq:g-coupl-sine},
when $\xi_0$ increases from
$\xi_0^{(0)}$ to $\xi_0^{(+)}$,
the  parameter $\xi_1$ changes from
$\xi_1^{(+)}$ to $\pi$. The equation for $\xi_1^{(+)}$ is 
\begin{align}
  \label{eq:xi1plus-sine}
  g_{+}(\xi_{1}^{(+)})=g_{-}(\xi_0^{(0)}).
\end{align}

For the parameters $q_0$ and $q_{+}$,
it can be inferred from the above results
that $q_{+}$ monotonically decays
from $q_1^{(+)}$ to zero with
$q_0$ growing from zero to $q_0^{(+)}$. 
The values of $q_0^{(+)}$ and $q_1^{(+)}$
can be computed from the relations 
\begin{align}
\label{eq:q01plus-sine}
  q_0^{(+)}=G_{0}^{(\sine)}(\xi_0^{(+)}),
\quad
  q_1^{(+)}=G_{+}^{(\sine)}(\xi_1^{(+)}), 
\end{align}
where $\xi_0^{(+)}$ and $\xi_1^{(+)}$
are given in Eq.~\eqref{eq:xi0plus0-sine}
and Eq.~\eqref{eq:xi1plus-sine}, respectively.

So, we arrive at the conclusion that 
the parameterized mapping~\eqref{eq:Psi-plus-sine}
behaves just like the half-period mapping
for the square wave voltage~\eqref{eq:Psi-plus-sqr}.
In addition, numerical calculations show that
using the formulas~\eqref{eq:Psi-explicit-sqr}
as a linear approximation for
the parameterized mappings $\Psi_{+}$ and $\Psi_{-}$ 
does not introduce any noticeable errors.
[The relative error was typically below $10^{-5}$.]  

It immediately follows that the cycle condition~\eqref{eq:cycle-sqr}
and Eq.~\eqref{eq:q-st-sqr} remain applicable for the sine-wave
voltages. 
However, the switching time parameters $\xi_{+}$ and $\xi_{-}$
can only be found numerically by solving Eq.~\eqref{eq:tau-pm}
where the relations~\eqref{eq:R-sine} and~\eqref{eq:q0-sine} 
are used to define the function~\eqref{eq:F} and to evaluate 
the values of $\xi_0$ for $q_{\pm}^{(\st)}$, respectively. 

For the most part, qualitative predictions of the model
for the sine-wave voltage remain unaltered
as compared to the case of the square wave voltage.
The cycles are found to be unstable and 
the condition~\eqref{eq:unstable-fixed} 
determines the cycle region.
Despite the critical period is longer than $T_c$
given by Eq.~\eqref{eq:T-c-sqr}, its dependence
on the voltage and polar anchoring parameters 
is qualitatively the same.

Similar to Eq.~\eqref{eq:u-st-approx-sqr},
the $T$-dependence of the parameters $u_{\pm}^{(\st)}$
turned out to be approximately linear,
whereas the switching time parameters
$\xi_{\pm}$ show a weak dependence on the frequency.
The effects of polarity are analogous to those discussed at the end of 
Sec.~\ref{subsec:square-wave-voltage}.

Thus, for all the wave forms under consideration,
the fixed points of the mapping~\eqref{eq:Psi-minus}
are repelling and the resulting instability of cycles
is the characteristic feature of the model~\eqref{eq:dyn-u}. 
Strictly speaking, it means that initially small deviations from
a cycle will grow in time.

In Appendix~\ref{sec:double-well-pot} our analysis is extended
to the generalized model with the double-well potential
which is a continuous deformation of Eq.~\eqref{eq:dyn-u}.
Equivalently, it can be regarded as a modification of
the Rapini-Papoular potential.
The key results can be summarized as follows
\renewcommand{\theenumi}{\alph{enumi}}
\renewcommand{\labelenumi}{(\theenumi)}
\begin{enumerate}
\item 
There are two limiting frequencies, $f_1$ and $f_2$,
so that the cycles are unstable at $f_2<f_w<f_1$
(see Eq.~\eqref{eq:dw-T12-sqr} and Eq.~\eqref{eq:dw-Ti-approx-sqr}).
\item 
There are no unstable cycles until
the separation between the minima of the potential
is sufficiently large (see the condition~\eqref{eq:dw-case}).
\item 
In addition to unstable cycle, there is the branch of stable cycles
in the low frequency region below the critical frequency $f_c$,
$f_w<f_c$ 
(see Eqs.~\eqref{eq:dw-Tc-sqr}--\eqref{eq:dw-Tc-approx-sqr}).
\item 
At $w_p\ne 0$, the low frequency region is separated from
the frequency interval of unstable cycles by the gap
of the width $f_2-f_c$. 
\end{enumerate}

These results suggest that 
the Rapini-Papoular anchoring potential
breaks down at large angular deviations of the director
rendering the cycles unstable.
For similar reasons, 
modifications of the potential were considered
in very recent studies
dealing with the dynamics of a pitch jump in
cholesteric liquid crystal
cells~\cite{Belyak:pre:2005,Belyak:jpcm:2006}.

%%%%%%%%%%%%%%%%%%%%%%%%%%%%%%%%%%%%
\begin{table}[htbp]
%\centering
\begin{ruledtabular}
  \begin{tabular}{lcccc}
  &
& $P_{\spn}$
& Tilt angle
& Viscosity
\\  
Mixture name 
& Phase sequence 
&  (nC/cm$^2$) 
&  $\theta$ (deg)
&  $\gamma$ (Pa s)
\\
\hline
FLC-497  
& Cr \myarrow{4} Sm-\chiC \myarrow{57} Sm-\chiA\myarrow{76} Iso
& 95
& 27
& 0.11
\\
FLC-510  
& Cr \myarrow{2} Sm-\chiC \myarrow{71} Sm-\chiA\myarrow{99} Iso
&  98
&  31
&  0.18
\\
\end{tabular}
\end{ruledtabular}
  \caption{Parameters of the FLC mixtures
measured in the Sm-\chiC\  phase at the temperature 23\degc.}
  \label{tab:flc-params}
\end{table}

%%%%%%%%%%%%%%%%%
\section{Experiment}
\label{sec:experim}
%%%%%%%%%%%%%%%%%

In this section we present the experimental
results on frequency dependence of the
threshold voltages measured in asymmetric 
SSFLC cells. It is found that the asymmetry induced shift
of the hysteresis loop increases with the driving frequency.
We apply the theoretical results of Sec.~\ref{sec:switch-dyn}
to model the process of switching within the cells
and to estimate the polar anchoring parameter
from the experimental data.

%%%%%%%%%%%%%%%%%%%%
\subsection{Sample preparation}
\label{subsec:sample-preparation}
%%%%%%%%%%%%%%%%%%%%

In our experiments we used asymmetric FLC cells
where the FLC layer is sandwiched between two dissimilar substrates.
One of the substrates was covered with 
a photo-aligning substance~--~the azobenzene sulfric dye SD-1,
whereas the other one was simply washed in  N,N-dimethylformamide (DMF) and
covered with calibrated spacers.
By contrast, the substrates with  photoaligned films 
identical in anchoring properties were assembled
to form symmetric FLC cells. 
 
Following the procedure described in Ref.~\cite{Chig:lc:2002},
SD-1 was synthesized from corresponding 
benzidinedisulfonic acid using azo coupling.
The solution was spin-coated onto glass substrates with
indium-tin-oxide (ITO) electrodes at 800 rpm for 5 seconds
and, subsequently, at 3000 rpm for 30 seconds.
The solvent was evaporated on a hot plate at 140\degc\  for 10 minutes.

The surface of the coated film was illuminated with linearly polarized
UV light using a super-high-pressure Hg lamp through an interference
filter at the wavelength 365~nm and a polarizing filter. 
The intensity of light irradiated normally on the
film surface during 30 minutes was 6~mW/cm$^2$.

We used two different pitch-compensated liquid crystal mixtures:
FLC-497 and FLC-510 (from P. N. Lebedev Physical Institute of
Russian Academy of Sciences) as materials for the FLC layer.
The mixtures were injected into the cells in an isotropic phase by capillary action
at T=85\degc\  and T=100\degc\ 
for FLC-497 and FLC-510, respectively.
In these mixtures, the FLC helix is unwound
due to compensation by two chiral dopants 
with opposite sense of chirality
(opposite sign of handedness)
but the same sign of the spontaneous polarization~\cite{Pozhid:lc:1989}.
The parameters of the mixtures are listed in Table~\ref{tab:flc-params}.

\begin{figure*}[!tbh]
%\vskip5mm
\centering
\resizebox{155mm}{!}{\includegraphics*{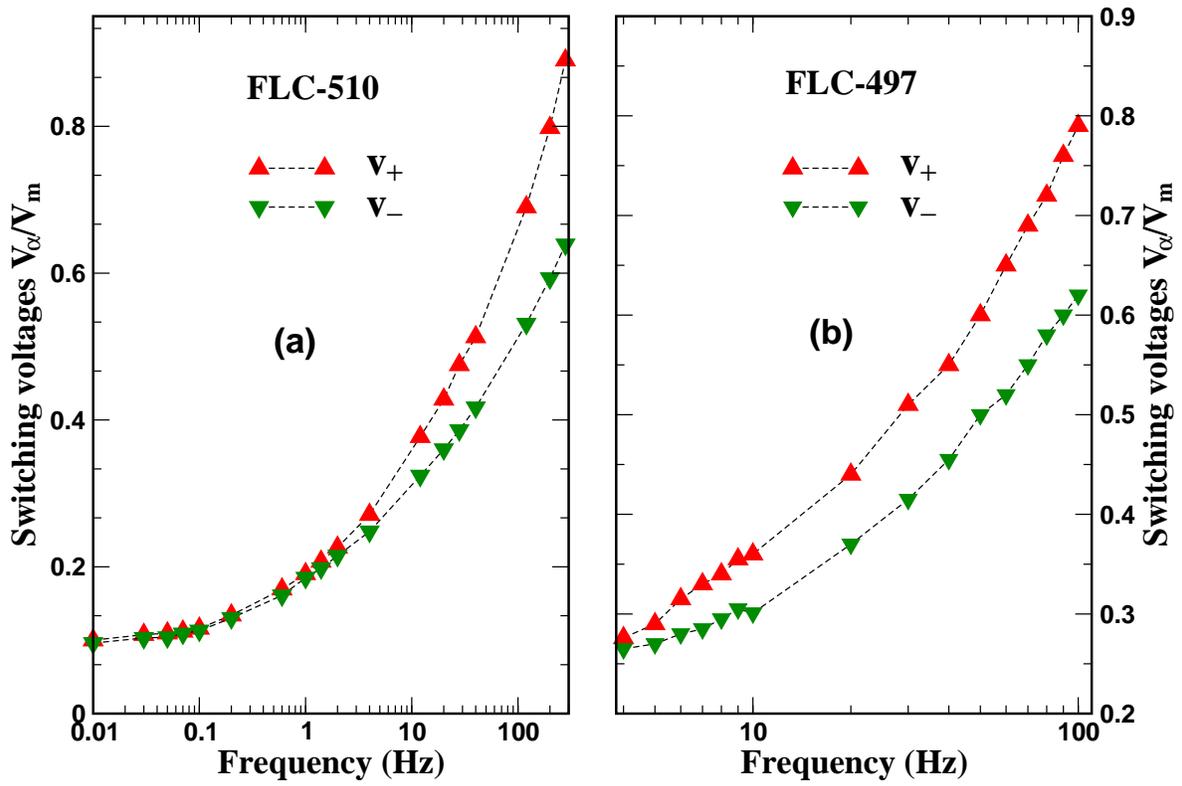}}
\caption{%
Experimental curves for
switching voltage parameters, $v_{\pm}=\pm V_{\pm}/V_{\mathrm m}$,
measured as a function of frequency
in the cells filled with (a)~FLC-510 and (b)~FLC-497.
}
\label{fig:sw-vpm-exp}
\end{figure*}

\begin{figure*}[!tbh]
%\vskip5mm
\centering
\resizebox{150mm}{!}{\includegraphics*{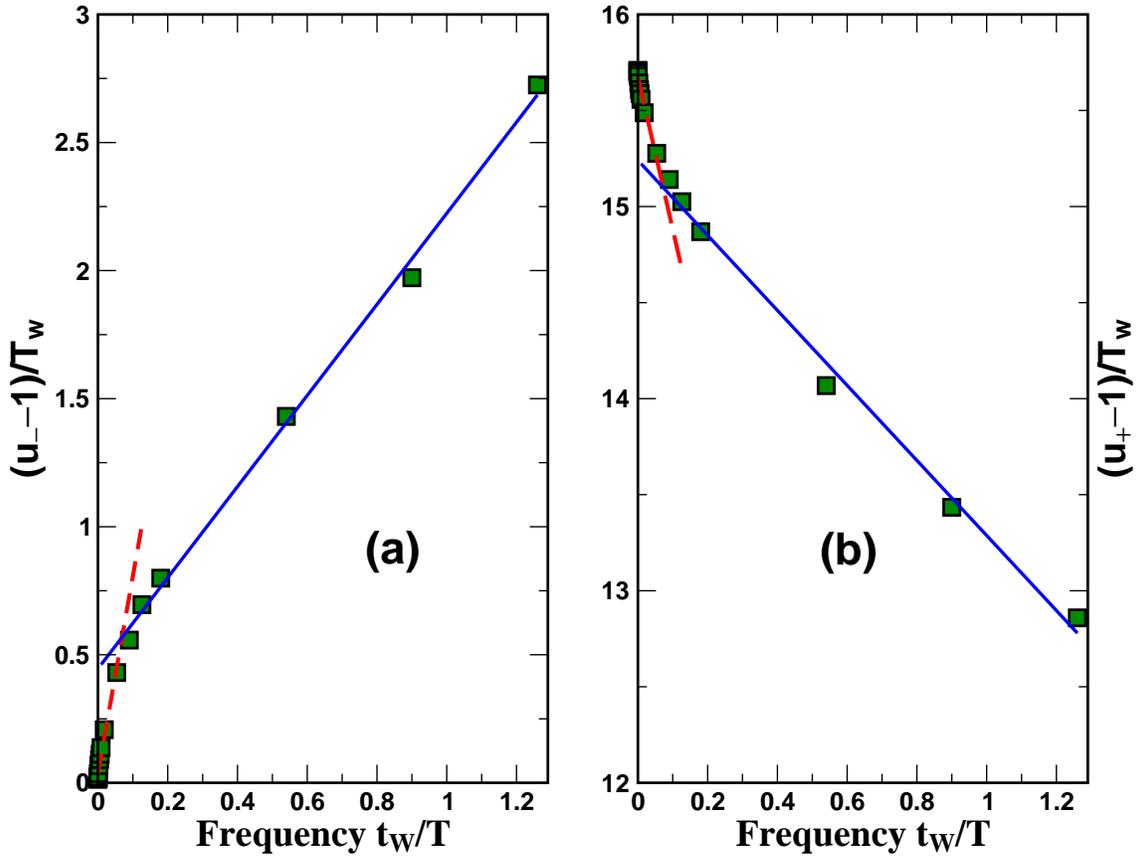}}
\caption{%
Frequency dependence of parameters (a)~$(u_{-}-1)/T_{w}$
and (b)~$(u_{+}-1)/T_{w}$
computed from the experimental data for the FLC-510 cell
using the model~\eqref{eq:dyn-u-approx} with $r_e=50$ and $w_p=-0.65$.
(The parameter $u_{\pm}$ determines the polarization of the up (down) state
$P_{\pm}=\pm P_{\spn}\tanh u_{\pm}$).
Piecewise linear approximation
of the results (solid lines) defines the ranges of high and low
frequencies with the boundary frequency about $0.07/t_W\approx 16$~Hz.
}
\label{fig:ang-varib-510}
\end{figure*}

 \begin{figure*}[!tbh]
%\vskip5mm
 \centering
 \resizebox{150mm}{!}{\includegraphics*{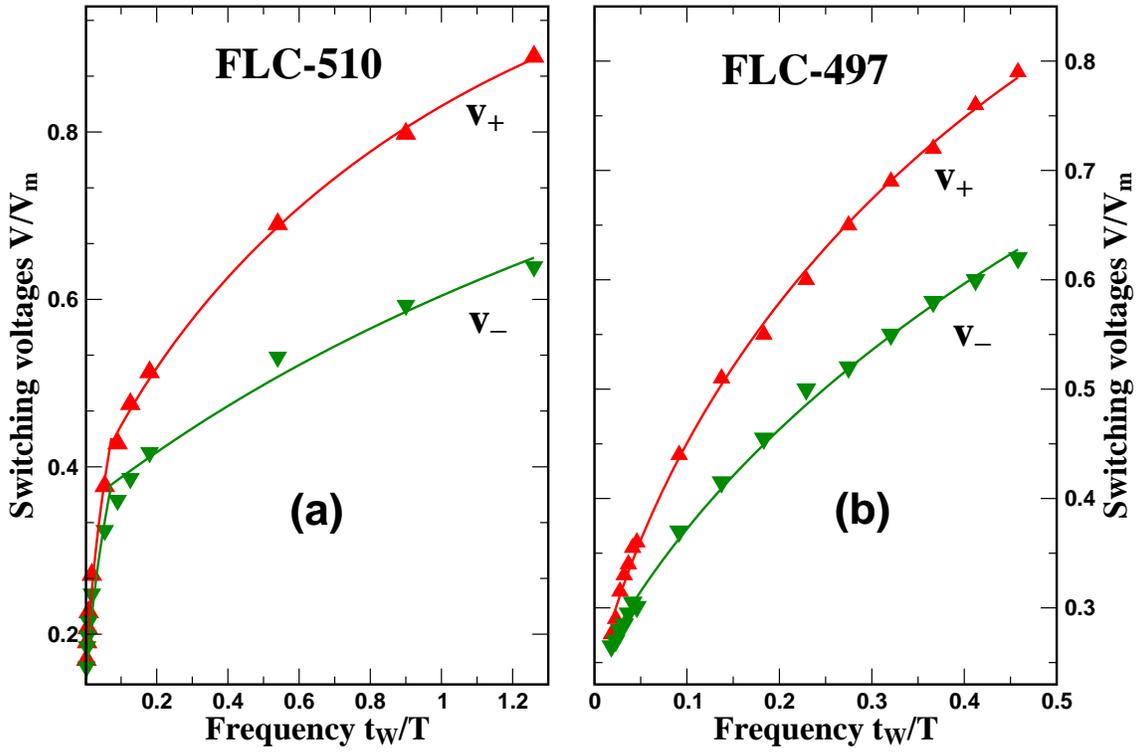}}
 \caption{%
 Switching voltage parameter computed 
as a function of frequency (solid lines)
using linearly approximated frequency dependence of $(u_{\pm}-1)f_w$
for (a)~the FLC-510 cell and (b)~the FLC-497 cell.
 }
 \label{fig:sw-vpm-freq}
 \end{figure*}

\begin{figure*}[!tbh]
%\vskip5mm
\centering
\resizebox{150mm}{!}{\includegraphics*{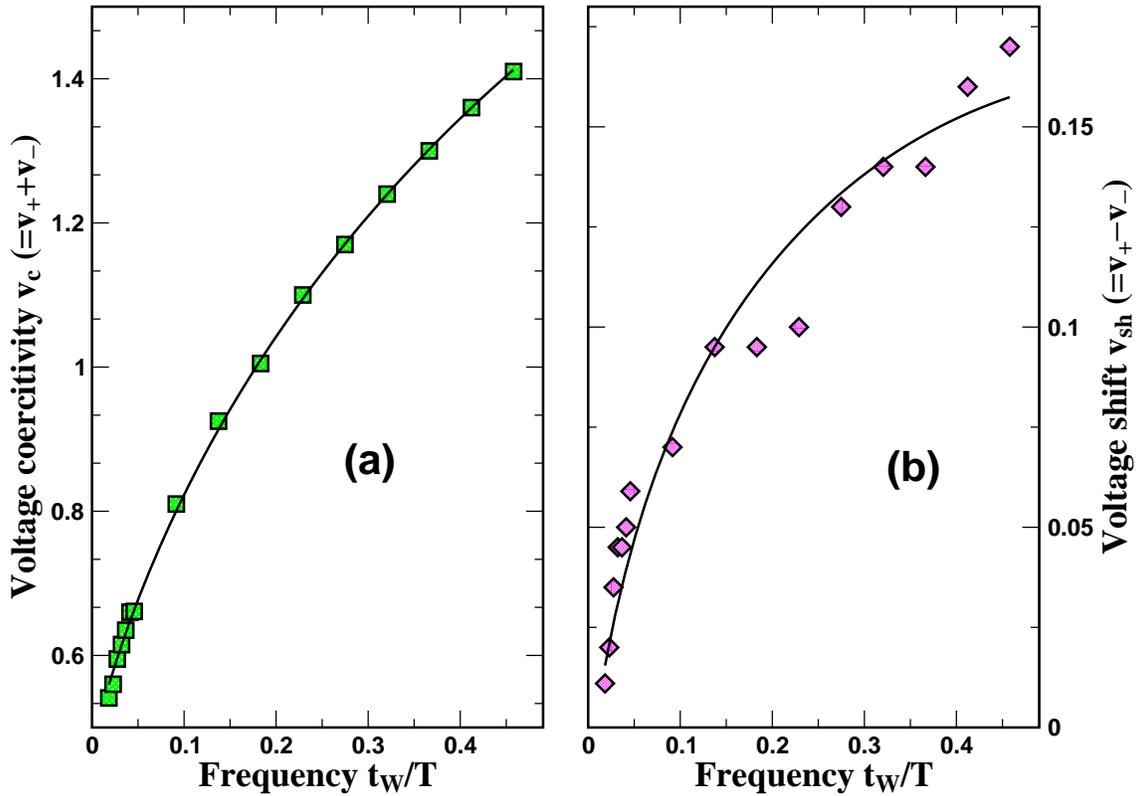}}
\caption{%
(a)~Voltage coercitivity and 
(b)~shift of the hysteresis loop as a function of frequency
for the FLC-497 cell.
Solid lines represent the results 
of modelling at $r_e=80$ and
$w_p=-0.78$ using the approximate relations:
$(u_{-}-1)f_w\approx 0.15+4.7 f_w$
and $(u_{+}-1)f_w\approx 15.6-4.24 f_w$.
}
\label{fig:coer-shift-497}
\end{figure*}

%%%%%%%%%%%%%%%%%%%%%%%%%%
\subsection{Experimental results and modelling}
\label{subsec:theory-vers-exper}
%%%%%%%%%%%%%%%%%%%%%%%%%%

Measurements of the hysteresis loops 
were performed using the standard electro-optical set-up 
composed of a He-Ne laser, a Hewlett
Packard Infinum digital oscilloscope and
a generator of triangular pulses. We also used
a rotating table for adjusting the angular
position of the FLC cell placed between crossed polarizers. 
The voltage amplitude of the generator, $V_{\mathrm m}$,
used in all experiments was 10~V and
the driving voltage frequency can be varied in the range from
10$^{-4}$ to 10$^3$~Hz. 

Figure~\ref{fig:hysteresis} schematically represents
the form of a typical hysteresis loop measured in our experiments.
The threshold voltages, $V_{\pm}$, and the corresponding values of
the switching voltage parameters, $v_{\pm}=\pm V_{\pm}/V_{\mathrm m}$,
then can be extracted
from the experimental data for the electro-optical hysteresis loops. 
For asymmetric cells filled with FLC-510 and FLC-497,
the results shown in Fig.~\ref{fig:sw-vpm-exp}
clearly indicate the difference between the voltage parameters
for switching up and down, $v_{+}$ and $v_{-}$.
Therefore, the hysteresis loops are shifted.
By contrast, for symmetric cells, 
the shift, $v_{+}-v_{-}$, is found to be vanishing
within the limits of experimental error.

The curves presented in Fig.~\ref{fig:sw-vpm-exp}
were measured in the cells where
the thickness of the FLC layer, $d$, and of the aligning
film, $d_A$, was 5~\mum\  and 12~nm, respectively.
In our previous papers~\cite{Chig:pre:2003,Kis:pre2:2005,Chig:pre:2005}, 
we studied anchoring properties of the azo-dye films
and our experimental technique
can be applied to measure the strength of anchoring $W$ 
in FLC cells.
It was estimated to be about
$2\times 10^{-4}$~J~m$^{-2}$
and $1.2\times 10^{-4}$~J~m$^{-2}$
in the cells filled with
FLC-510 and FLC-497, respectively. 

The characteristic times $t_E$ and $t_W$
given in Eq.~\eqref{eq:time-scales}
can now be estimated as follows:
$t_E\approx 0.9\times 10^{-4}$~s
and $t_W\approx 4.5\times 10^{-3}$~s
for the FLC-510 cell;
$t_E\approx 0.58\times 10^{-4}$~s
and $t_W\approx 4.58\times 10^{-3}$~s
for the FLC-497 cell. 
The corresponding values of the driving voltage parameter are:
$r_e\approx 50$ and $r_e\approx 79$.

The above estimates define the parameters 
that enter the dynamical equation~\eqref{eq:dyn-phi}
of the model discussed in Sec.~\ref{subsec:energy}.
In this model the director field~\eqref{eq:director}
is assumed to be uniform across the cell.
Such assumption can be a reasonable approximation
in the high field regime of switching.
This is the case in which
the characteristic length $\xi_E=(K/P_{\spn} E_0)^{1/2}$,
where $K$ is the effective elastic constant,
is shorter than the cell thickness, $\xi_E<d$.
From
the estimate $\xi_E\approx 0.07$~\mum\ $\ll d=5.0$~\mum\
obtained using a typical value of the elastic constant,
$K\approx 10^{-11}$~J~m$^{-1}$,
we may conclude that our measurements were
performed in the high field regime and the uniform model
can be used to interpret the experimental results.

We can also estimate the anchoring extrapolation length 
$\xi_W=K/W\approx 0.05-0.08$~\mum, so that 
the characteristic lengths $\xi_E$ and $\xi_W$ are of the same order.
Therefore, the anchoring conditions at the boundary surfaces
described by the anchoring energy
could play an important part in the switching dynamics.

In Sec.~\ref{subsec:energy},
we found the expression for
the electric field inside the FLC layer~\eqref{eq:E_y-uni}
which differ from the external electric field
due to the presence of the insulating aligning film.
This effect crucially depends on the value of the dimensionless
parameter $\eta$ defined in Eq.~\eqref{eq:E_0-D}
that, in our case, can be estimated 
by assuming that
$d_2=0$, $d_1=d_A=12$~nm and $\varepsilon_1=\varepsilon_A\approx 7.5$
is the dielectric constant of the SD-1 layer.
For both FLC-510 and FLC-497 cells, 
we obtain small values of $\eta$ that are below $5\times 10^{-4}$.
As a result, even a rough estimate for the depolarizing voltage
$V_{\mathrm{dep}}\approx P_{\spn}
d_A/(\varepsilon_A\varepsilon_0)\approx 0.16$~V
derived from Eq.~\eqref{eq:E_y-uni} yields the voltages
an order of magnitude lower than 
typical values of the voltage shift measured in our experiments. 
In addition, a more accurate analysis
of the contributions to
the effective electrostatic free energy~\eqref{eq:f_eff}
clearly shows that the voltage drop across the aligning layer
cannot be responsible for asymmetry of the switching voltage 
thresholds, $V_{+}\ne V_{-}$.

The method developed in Sec.~\ref{sec:switch-dyn}
can now be applied to model switching dynamics
of the FLC cells using the experimental results.
To this end we calculated the parameters $u_{\pm}$
characterizing the polarization of the up and down states 
from the data on frequency
dependence of the switching voltage parameters $v_{\pm}$.
This procedure involves solving Eq.~\eqref{eq:tau-pm}
and using the half-period mapping~\eqref{eq:Psi-plus-gen} 
for the triangular waveform function~\eqref{eq:triang-wv}.
The mapping $\Psi_{+}$ ($\Psi_{-}$)
is used to compute  the parameters $u_{\pm}$ for switching up (down) 
from the data on the switching voltage $v_{+}$ ($v_{-}$).

In general, the computed values of $u_{\pm}$ for switching up 
deviate from the corresponding results for switching down.
We used the polar anchoring parameter $w_p$ as a fitting parameter
to minimize the difference
and to obtain the results nearest to the cycle.
For the FLC-510 cell, the value of $w_p$ is found
to be about $-0.65$ and 
we present the results for the parameters $(u_{\pm}-1) f_w$ 
in Fig.~\ref{fig:ang-varib-510}. Referring to
Fig.~\ref{fig:ang-varib-510}, there are two frequency regions
where, similar to cycles, the frequency dependence of
$(u_{\pm}-1) f_w$ is nearly linear.
Using linear approximation for the parameters
$(u_{\pm}-1) f_w$ we estimated the boundary frequency
separating the low and high frequency regions at about $16$~Hz
and calculated the curves for the switching voltage parameters
shown in Fig.~\ref{fig:sw-vpm-freq}a as solid lines.

The switching voltage vs frequency curves computed along the same lines
for the FLC-497 cell are given in Fig.~\ref{fig:sw-vpm-freq}b.
In this case, the polar anchoring parameter is found to be about
$-0.78$. From Fig.~\ref{fig:coer-shift-497}a it can be seen that
the linear approximation in the high frequency region
reproduces the experimental data for the voltage coercitivity 
remarkably well. The scatter in the data and its scale
are more evident in the results for the voltage shift shown in
Fig.~\ref{fig:coer-shift-497}b.

Despite good agreement between  
the theoretical and the experimental results,
it is essential that the linearly fitted data, strictly speaking,
do not represent the cycle and the modelling was only performed
over the first period. Since the cycles are unstable,
evolution in time will have the deterioration effect on the agreement
between modelling and measurements.

%%%%%%%%%%%%%%%%%%%%%%%%%%%%%
\section{Discussion and conclusions}
\label{sec:discussion}
%%%%%%%%%%%%%%%%%%%%%%%%%%%%%

In this paper we have studied 
how the dynamical characteristics of switching in SSFLC cells 
are affected by asymmetry of dissimilar substrates
in anchoring properties.
The asymmetry effects are found to be dominated by the polar contribution
to the anchoring energy and manifest themselves in the frequency
dependent shift of the hysteresis loop.

Our theoretical considerations were primarily concerned with
the predictions of the uniform theory on the frequency dependence
of the switching voltage thresholds (the switching voltage
parameters).  It was assumed that 
the steady state regime of switching 
is determined by the cycles which are time-periodic
solutions to the dynamical equation.
So, we have developed the method to analyze the properties of cycles
depending on the driving frequency.
In this method the dynamical equation for the angular
variable~\eqref{eq:vos-u-var} 
is approximated to yield the half-period mappings in the parameterized
form. The cycles then studied in terms of the fixed points of
the composition of two half-period mappings.

By using this method we performed analysis for the cases
of square wave, sine-wave and triangular voltages.
It was found that the cycles are unstable and
can only be formed when the driving frequency is lower
than its critical value, $f<f_c=1/T_c$.
The critical frequency $f_c$ declines with the polar anchoring
parameter $w_p$ suppressing the cycles at $|w_p|=1$. 

The polar anchoring breaks the mirror symmetry relating
the processes of switching up and down 
leading to the difference in the magnitude of
the corresponding switching time and switching voltage parameters.
So, we have $\tau_{+}\ne \tau_{-}$ and $v_{+}\ne v_{-}$
at non-vanishing polar anchoring parameter, $w_p\ne 0$.
Therefore, the shift of the hysteresis loop, $v_{+}-v_{-}$,
results from the symmetry breaking effect
caused by the polar contribution to the anchoring potential~\eqref{eq:anchoring-uni}
characterized by the polar anchoring parameter~\eqref{eq:meta-stability}.

Our calculations indicate that, for cycles, dependence of the parameters
$u_{\pm}^{(\st)}/T$ and $\tau_{\pm}/T$ on the frequency is nearly
linear. Modelling of the switching dynamics
using the experimental data revealed similar behavior 
provided the frequency is not too low.
Since the conductivity of the FLC layer
has been neglected in the electrostatic model
 discussed in Sec.~\ref{subsec:energy}, we cannot expect
the model to give accurate results at low frequencies.

The expression for the effective electrostatic
potential~\eqref{eq:f_eff} shows that
there are no symmetry breaking contributions
due to the voltage drop across the insulating aligning films
and the dielectric anisotropy of the FLC layer.
So, they cannot be responsible for the asymmetry effects under
consideration. Thus we may conclude that the anchoring energy
is the determining factor in this problem.
Of interest is the fact that similar conclusion 
underlines the method proposed in~\cite{Ulrich:lc:1997} 
to control the process of switching. 
 
During the switching process the azimuthal angle
varies between zero and $\pi$. This involves large deviations of the director
from the easy axis, so that the switching dynamics 
can be sensitive to the shape of the anchoring potential.
It can be assumed that 
the instability of cycles is a characteristic of 
the model with the potential taken in
the standard Rapini-Papoular form~\cite{Gennes:bk:1993}.
In the Appendix we have studied the generalized model
with the double-well potential and found that
the branch of stable cycles emerges even at small
deformations of the potential.
Note that the results derived in the Appendix
not only demonstrate the effectiveness of the analytical approach
but also can be placed in a more general physical context.

The  uniform theory cannot be directly applied to
more complicated cases with spatially inhomogeneous orientational 
structures involved. So, discussion of such problems as 
the soliton (kink) mechanism of 
switching~\cite{Clark:jap:1994,Pikin:pre:1995,Tsoy:pre:1999}
and dynamics in the chevron geometry~\cite{Copic:pre:2000,Hazel:lc:2004}
is beyond the scope of this paper.
Nevertheless, it is pertinent to note that, under certain circumstances,
the switching dynamics of inhomogeneous structures 
can be effectively reduced to  a uniform model
for relevant spatially independent variables 
using a trial solution in the first approximation.
 
\begin{acknowledgments}
This work is performed under the grant ITS/111/03.
 \end{acknowledgments}

\appendix

%%%%%%%%%%%%%%%%%%%%%%%%%%%%%%%%%%%%%%%%%%%%%
\section{Double-well potential: cycles
  in piecewise quadratic model}
\label{sec:double-well-pot}
%%%%%%%%%%%%%%%%%%%%%%%%%%%%%%%%%%%%%%%%%%%%%

In Sec.~\ref{sec:switch-dyn} we have studied the switching
dynamics by analyzing the dynamical model~\eqref{eq:dyn-u-approx}
for the angular variable $u$.
The assumption underlying the study is that the dynamics is governed
by the cycles which are periodic solutions to the dynamical equation
in the regime of complete switching.
Technically, 
we have used the analytical approach 
described in Sec.~\ref{subsec:half-period-mapping}
in which the cycles are related to 
the fixed points
of the composition of the half-period mappings~\eqref{eq:Psi-minus}.

One of the most important results presented in Sec.~\ref{subsec:square-wave-voltage}
and Sec.~\ref{subsec:sine-wave}
is the instability of the fixed points, so that 
there are no stable cycles in the model~\eqref{eq:dyn-u-approx}.
In this appendix, in order to clarify the mechanism of such
instability, we consider the generalized model 
with the governing equation 
\begin{equation}
  \label{eq:dw-dyn-u-approx}
  \pdr{u}{\tau}\equiv\dot{u} = -\pdr{U(u)}{u}+ r(\tau),
\quad
U(u)=U_{0}(u)-w_p u,
\end{equation}
where the anchoring energy potential $U_{0}$ is taken to be
a piecewise quadratic function of the following form
\begin{equation}
  \label{eq:dw-poten-gen}
  U_{0}(u)=
\begin{cases}
\epsilon_{-} u^2/2+(1+\epsilon_{-})(u+1/2),& u<-1,\\
-u^2/2, & -1\le u\le 1,\\
\epsilon_{+} u^2/2-(1+\epsilon_{+})(u-1/2),& u>1.
\end{cases}
\end{equation}
Clearly, our original model~\eqref{eq:dyn-u-approx}
can be regarded as the limiting case of the modified model~\eqref{eq:dw-dyn-u-approx}
in which the parameters $\epsilon_{\pm}$
are equal to zero, $\epsilon_{\pm}=0$.

%%%%%%%%%%%%%%%%%%%%%%%%%%%%%%%%%%%%

\begin{figure*}[!tbh]
%\vskip5mm
\centering
\resizebox{150mm}{!}{\includegraphics*{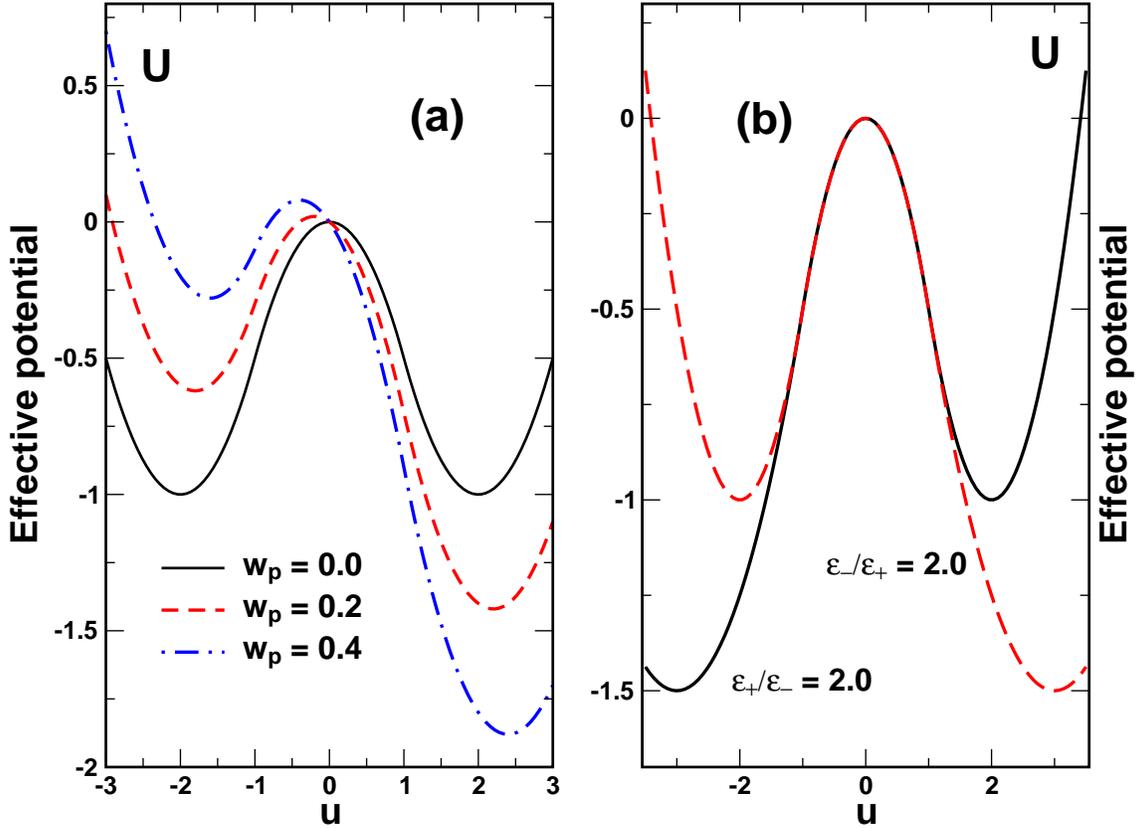}}
\caption{%
Double-well effective potential as a function of $u$. 
Two cases are shown:
(a)~$\epsilon_{\pm}=1$
at various values of 
the polar anchoring parameter $w_p$;
and
(b)~$\epsilon_{+}\ne\epsilon_{-}$
at $w_p=0$.
}
\label{fig:dw-poten}
\end{figure*}

\begin{figure*}[!tbh]
%\vskip5mm
\centering
\resizebox{150mm}{!}{\includegraphics*{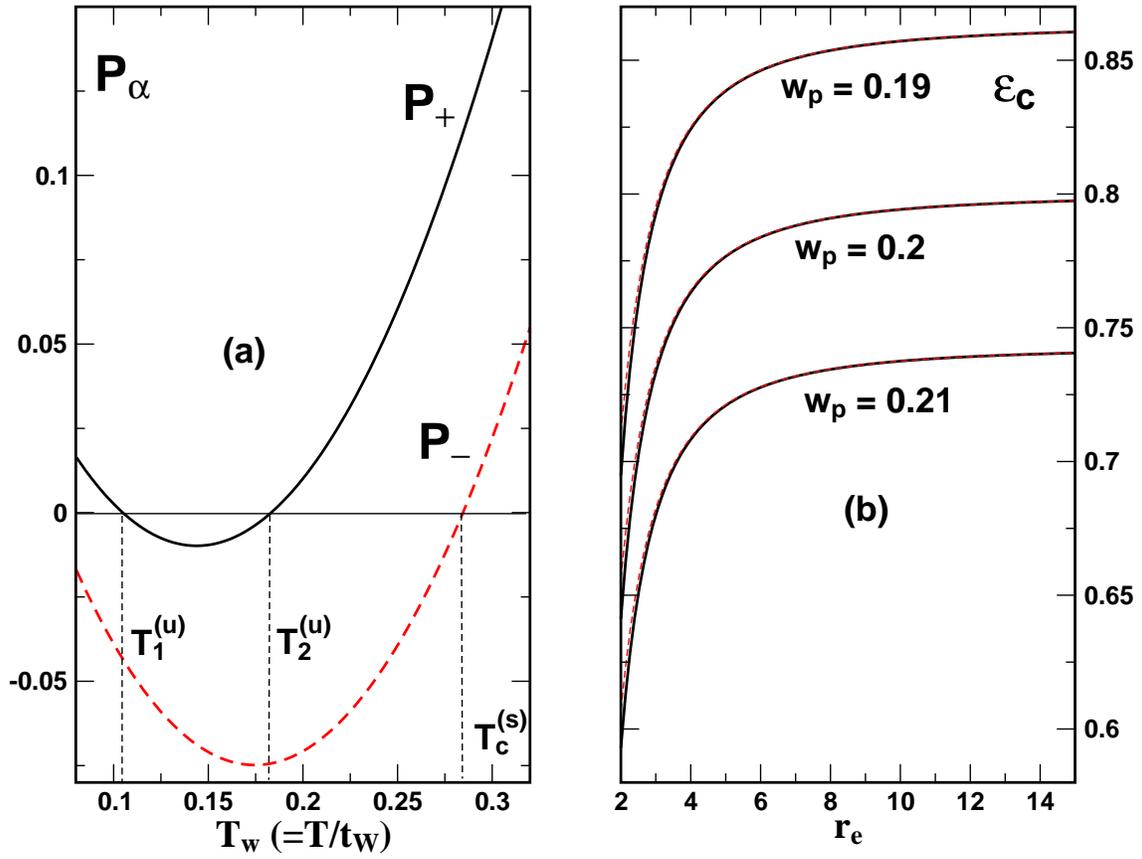}}
\caption{%
(a)~The functions $P_{\pm}$
versus the period $T_{w}$ computed at
$w_p=0.2$, $r_e=50$ and 
$\epsilon=0.5<\epsilon_c\approx 0.8$.
(b)~The critical value of $\epsilon$ as a function of the
voltage parameter
at various values of the polar anchoring parameter $w_{p}$.
The dashed curves are computed from the asymptotic 
formula~\eqref{eq:dw-epsc-approx-sqr}.
}
\label{fig:crit-epsilon-eps}
\end{figure*}

\begin{figure*}[!tbh]
%\vskip5mm
\centering
\resizebox{150mm}{!}{\includegraphics*{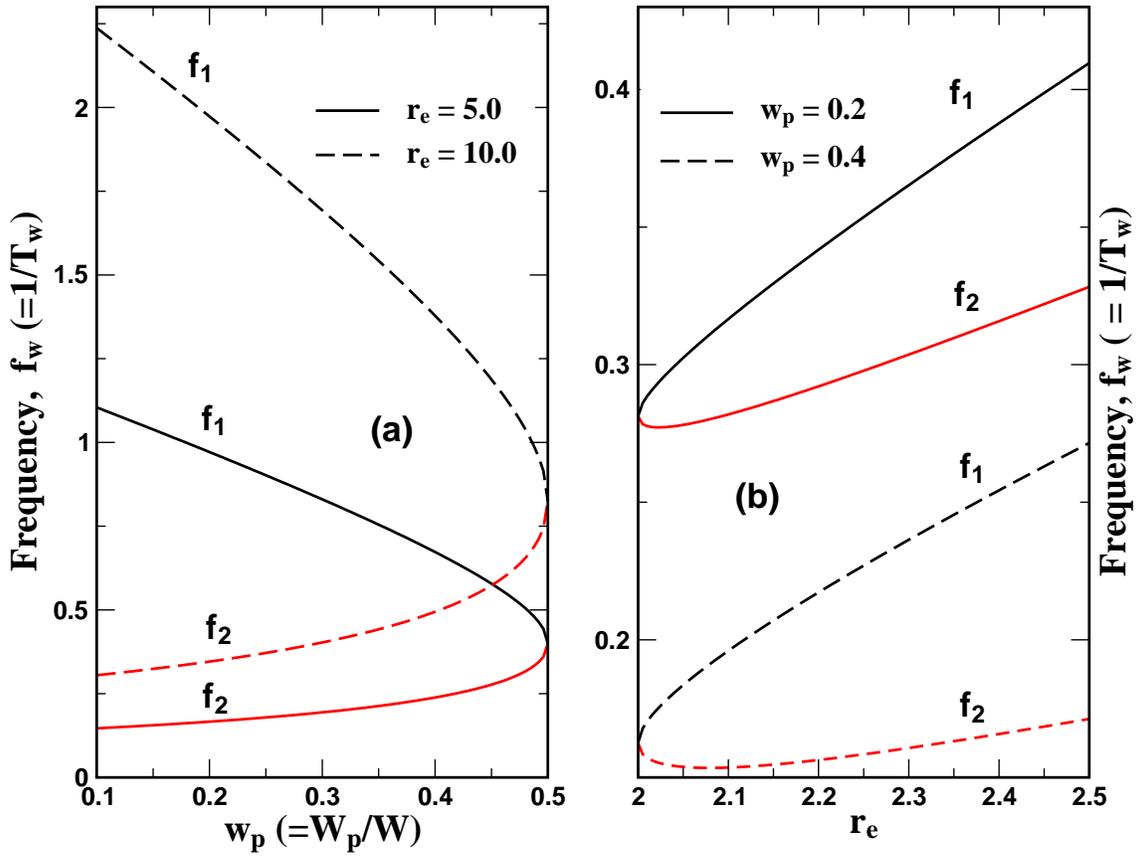}}
\caption{%
(a)~Region of unstable cycles,
$f_2<f_{w}<f_1$,
in the   $w_p$-$f_{w}$ plane.
At $r_e=5.0$ and $r_e=10.0$,
the curves for the limiting frequencies,
$f_1$ and $f_2$,
are shown as solid and dashed lines,
respectively.
(b)~The region,
$f_2<f_{w}<f_1$, in the   $r_e$-$f_{w}$ plane.
Solid and dashed lines represent
the curves for the limiting frequencies
computed at $w_p=0.2$ and $w_p=0.4$, 
respectively.
}
\label{fig:crit-unst-eps}
\end{figure*}

\begin{figure*}[!tbh]
%\vskip5mm
\centering
\resizebox{150mm}{!}{\includegraphics*{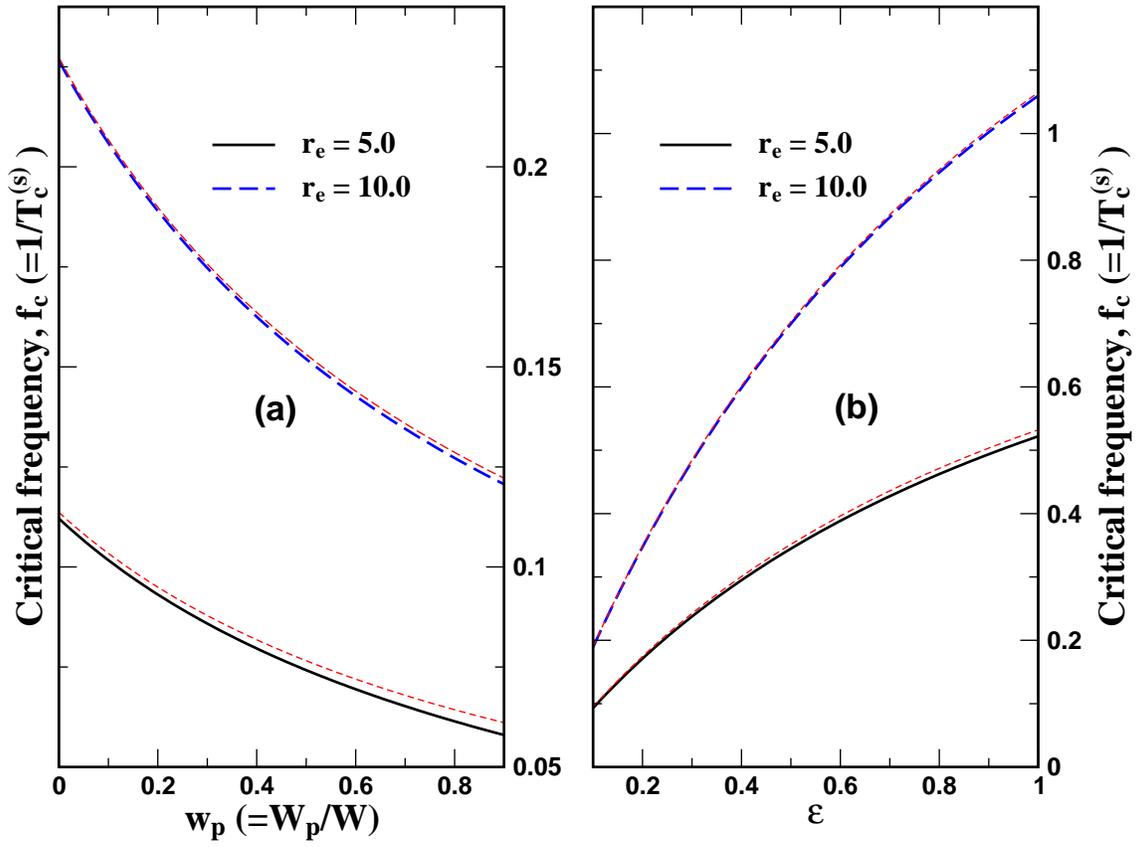}}
\caption{%
Critical frequency of stable cycles, 
$f_c=1/T_c^{(s)}$ [see Eq.~\eqref{eq:dw-Tc-sqr}], as a function 
of (a)~the polar anchoring  parameter
and (b) the parameter~$\epsilon$.
The asymptotic formula~\eqref{eq:dw-Tc-approx-sqr}
is used for computing the curves shown as
thin dashed lines.
}
\label{fig:crit-stbl-eps}
\end{figure*}

\begin{figure*}[!tbh]
%\vskip5mm
\centering
\resizebox{150mm}{!}{\includegraphics*{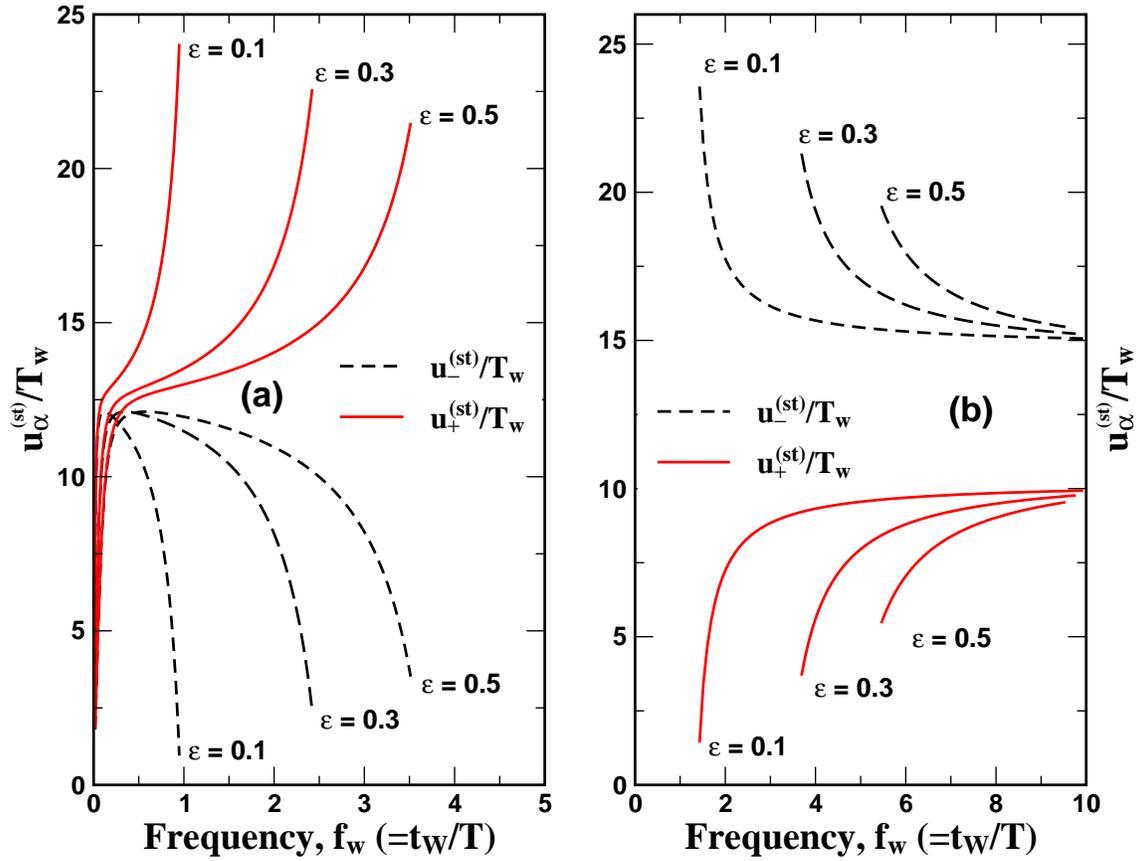}}
\caption{%
Frequency dependence of parameters
$u_{\pm}^{(\st)}/T_{w}$ for (a)~stable and (b)~unstable cycles
with $w_p=0.2$ and $r_e=50$
at various values of $\epsilon$.
}
\label{fig:u-st-eps}
\end{figure*}

\begin{figure*}[!tbh]
%\vskip5mm
\centering
\resizebox{150mm}{!}{\includegraphics*{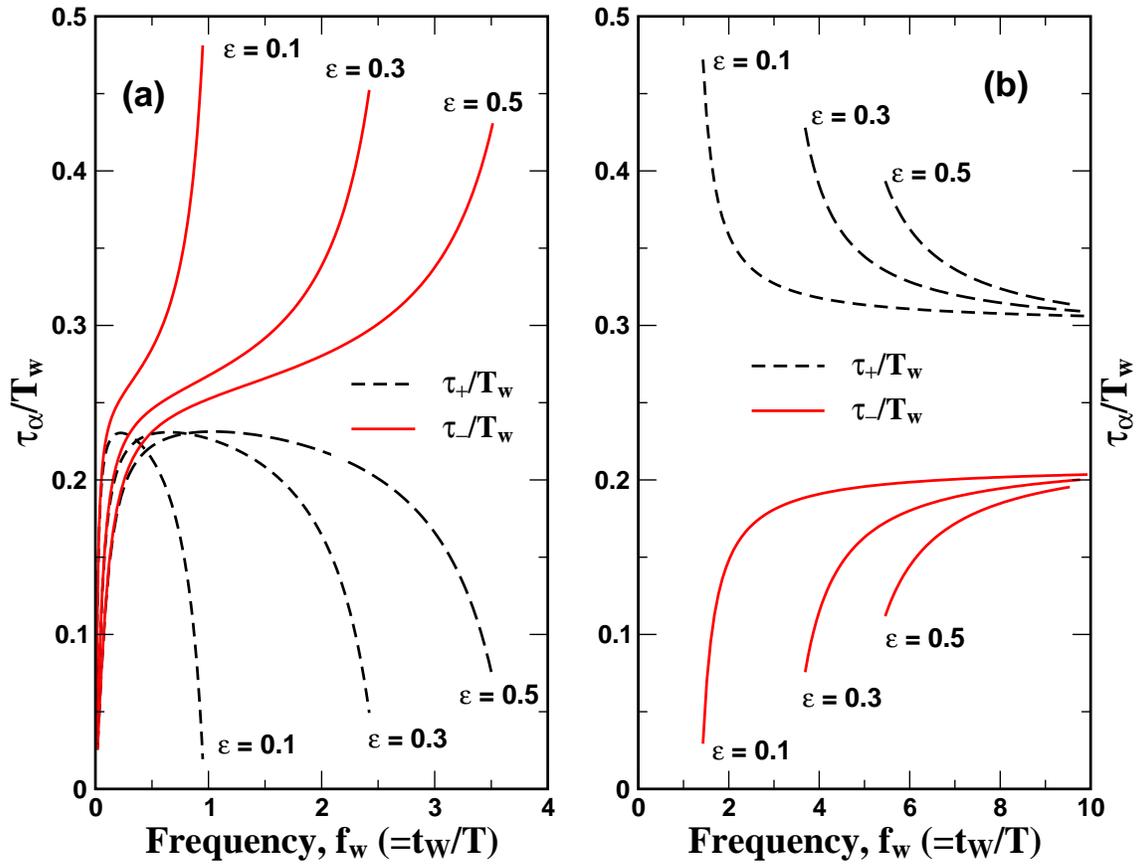}}
\caption{%
Frequency dependence of switching time parameters
$\tau_{\pm}/T_{w}$ for (a)~stable and (b)~unstable cycles
with $w_p=0.2$ and $r_e=50$
at various values of $\epsilon$.
}
\label{fig:tau-pm-eps}
\end{figure*}
%%%%%%%%%%%%%%%%%%%%%%%%%%%%%%%%%%

When $\epsilon_{\pm}>0$, $U$ is a double-well potential 
that have two local minima representing the equilibrium up and down states 
with $u=u_{\mn}^{(+)}$ and $u=u_{\mn}^{(-)}$, respectively. 
These are separated by the maximum
located at $u=u_{\mx}$ and are given by
\begin{align}
&
\label{eq:dw-U-min}
  u_{\mn}^{(\pm)}=\pm (1+\epsilon_{\pm}\pm
  w_p)/\epsilon_{\pm},\quad
2 U_{\mn}^{(\pm)}=1+\epsilon_{\pm}-
(1+\epsilon_{\pm}\pm w_p)^2/\epsilon_{\pm},
\\
&
  \label{eq:dw-U-max}
  u_{\mx}=-w_p,\quad
  U_{\mx}=w_p^2/2.
\end{align}
From Eq.~\eqref{eq:dw-U-min} the minima are energetically equivalent
and the states characterized by $u_{\mn}^{(+)}$ and
$u_{\mn}^{(-)}=-u_{\mn}^{(+)}$ are stable  at
$\epsilon_{+}=\epsilon_{-}$ and $w_p=0$.
This is no longer the case when $w_p\ne 0$.
As is illustrated in Fig.~\ref{fig:dw-poten}a,
similar to the anchoring potential~\eqref{eq:anchoring-uni}
[see Fig.~\ref{fig:poten-phi}],
the down state is metastable at $w_p>0$ and the energy barrier
disappears when the polar anchoring parameter approaches unity.

Referring to Fig.~\ref{fig:dw-poten}b, the asymmetry effects
can also be caused by the difference between $\epsilon_{+}$
and $\epsilon_{-}$. But, for brevity, we shall restrict
ourselves to the case specified by the conditions 
\begin{equation}
  \label{eq:dw-cond0}
  \epsilon_{\pm}=\epsilon>0,
\quad
|w_p|<1.
\end{equation}

In the limit of vanishing $\epsilon$, $\epsilon\to 0$,
the parameters $|u_{\mn}^{(\pm)}|$ become infinitely large
and the minima shift at infinity in opposite directions.
We may anticipate that this divergence
renders the cycles unstable and
our study of the model with
the double-well potential~\eqref{eq:dw-poten-gen}
will confirm this hypothesis.

We perform our analysis following the general procedure described in
Sec.~\ref{subsec:half-period-mapping}.
For the dynamical equation~\eqref{eq:dw-dyn-u-approx},
the structure of the solutions is defined by Eq.~\eqref{eq:u-sol},
where the functions $F_0$ and $F_{+}$ need to be changed as follows
\begin{subequations}
\label{eq:dw-F-mod}
\begin{align}
\label{eq:dw-F0}
F_{0}(\tau)=&-u_0\exp[-\epsilon\tau]+
(w_p-1-\epsilon)(1-\exp[-\epsilon\tau])/\epsilon
+\exp[-\epsilon\tau]\,R_{\epsilon}(\tau),
\\
\label{eq:dw-Fplus}
F_{+}(\tau)=&\exp[-\epsilon(\tau-\tau_{1})]
+(w_p+1+\epsilon)
\bigl(
1-\exp[-\epsilon(\tau-\tau_{1})]
\bigr)/\epsilon
+
\nonumber
\\
&
\exp[-\epsilon\tau]\,
\bigl(
R_{\epsilon}(\tau)-R_{\epsilon}(\tau_{1})
\bigr),
\\
  \label{eq:dw-R0}
  R_{\epsilon}(\tau)=&\int_{0}^{\tau} \exp(\epsilon\tau') r(\tau')\dd\tau'.   
\end{align}
\end{subequations}
The corresponding modification of
the functions $G_0$ and $G_{+}$ 
that enter the half-period mapping~\eqref{eq:Psi-plus-gen} is given by
\begin{subequations}
\label{eq:dw-G-mod}
\begin{align}
\label{eq:dw-G0}
G_{0}(\tau_0)=&
(w_p-1)
\bigl(
\exp[\epsilon\tau_0]-1
\bigr)/\epsilon
+R_{\epsilon}(\tau_0),
\\
\label{eq:dw-Gplus}
G_{+}(\tau_1)=&\exp[-\epsilon T_{w}/2]
\Bigl\{
(w_p+1)
\bigl(
\exp[\epsilon T_{w}/2]-
\exp[\epsilon\tau_1]
\bigr)/\epsilon
+
\nonumber
\\
&
R_{\epsilon}(T_{w}/2)-R_{\epsilon}(\tau_{1})
\Bigr\}.  
\end{align}
\end{subequations}
Note that the coupling equation~\eqref{eq:g-coupl}
as well as the function $R$ defined in Eq.~\eqref{eq:R0}
remain untouched.

%%%%%%%%%%%%%%%%%%%%%%%%%
\subsection{Square wave voltage}
\label{subsec:sq-wave-voltage-eps}
%%%%%%%%%%%%%%%%%%%%%%%%%

For the square wave voltage,
Eqs.~\eqref{eq:dw-G0} and~\eqref{eq:dw-Gplus}
take the simplified form
\begin{subequations}
\label{eq:dw-G-sqr}
\begin{align}
\label{eq:dw-G0-sqr}
G_{0}(\tau_0)=&
(w_p+r_e-1)
\bigl(
\exp[\epsilon\tau_0]-1
\bigr)/\epsilon,
\\
\label{eq:dw-Gplus-sqr}
G_{+}(\tau_1)=&
(w_p+r_e+1)
\bigl(
1-
\exp[\epsilon(\tau_1-T_{w}/2)]
\bigr)/\epsilon.   
\end{align}
\end{subequations}
These expressions can now be combined with
the coupling equation~\eqref{eq:tau01-sqr}
to yield the linear half-period mapping
of the form~\eqref{eq:Psi-explicit-sqr}.
But, in place of Eq.~\eqref{eq:alpha-sqr},
we have
\begin{subequations}
\label{eq:dw-q-sqr}
\begin{align}
&
  \label{eq:dw-q1-sqr}
q_1^{(\pm)}=
(\pm w_p+r_e+1)
\bigl(1-\exp[\epsilon(\gamma_{\pm}-T_{w}/2)]
\bigr)/\epsilon,
\\
&
\label{eq:dw-q0-sqr}
q_0^{(\pm)}=
(\pm w_p+r_e-1)
\bigl(
\exp[\epsilon(T_{w}/2-\gamma_{\pm})]-1
\bigr)/\epsilon,  
\end{align}
\end{subequations}
where $\gamma_{\pm}$ is given in Eq.~\eqref{eq:tau01-sqr}.

Following the same line of reasoning as in
Sec.~\ref{subsec:square-wave-voltage},
we derive the conditions of complete switching
for the voltage parameter 
and the frequency 
(see Eq.~\eqref{eq:switch-r-sqr} and 
Eq.~\eqref{eq:T-min-sqr}, respectively).
We also find that the cycles may only develop 
provided the cycle condition~\eqref{eq:cycle-sqr} is met.

This condition is expressed in terms of the parameters $P_{\pm}$.
Substituting the relations~\eqref{eq:dw-q-sqr} 
into Eq.~\eqref{eq:cycle-sqr}
yields these parameters
\begin{align}
&
  \label{eq:dw-Ppm-sqr}
\epsilon z P_{\pm}=
a_{\pm} z^2 - 2 r_e z +b_{\pm},
\quad
z\equiv \exp(-\epsilon T_{w}/2),
\\
&
\label{eq:dw-abpm-sqr}
a_{\pm}=
(\mp w_p+r_e+1)[\beta_{\mp}]^{\epsilon},
\quad
b_{\pm}=
(\pm w_p+r_e-1)[\beta_{\pm}]^{-\epsilon},
\end{align}
where $\beta_{\pm}$ is defined in Eq.~\eqref{eq:tau01-sqr}.
It remains to recall that the cycles may build up only if $P_{+}P_{-}>0$.

From Eq.~\eqref{eq:dw-Ppm-sqr} it is clear that 
the sign of the parameters $P_{\pm}$ is determined by the quadratic
functions of the period dependent variable $z$ that lies in the interval
\begin{equation}
  \label{eq:z-min-sqr}
  0<z<z_{\mx},\quad
z_{\mx}=\exp(-\epsilon T_{\mn}/2),
\end{equation}
where $T_{\mn}$ is given by Eq.~\eqref{eq:T-min-sqr}.

The discriminants of the quadratic polynomials~\eqref{eq:dw-Ppm-sqr}
\begin{equation}
  \label{eq:dw-Dpm-sqr}
  D_{\pm}=r_e^2-
\bigl(
r_e^2-(1\mp w_p)^2
\bigr)
\biggl[
\frac{r_e^2-(1\mp w_p)^2}{r_e^2-(1\pm w_p)^2}
\biggl]^{\epsilon}
\end{equation}
are both positive when the parameter $\epsilon$
is below its critical value given by
\begin{equation}
  \label{eq:dw-epsc-sqr}
  \epsilon_c=
\ln\frac{r_e^2}{r_e^2-(1- |w_p|)^2}
\biggl[
\ln\frac{r_e^2-(1- |w_p|)^2}{r_e^2-(1+ |w_p|)^2}
\biggr]^{-1}.
\end{equation}
In the region of high voltage parameters,
the asymptotic formula for $\epsilon_c$ is
\begin{equation}
  \label{eq:dw-epsc-approx-sqr}
  \epsilon_c\approx
|w_p|^{-1} (1-|w_p|)^2\bigl(
2-(1+|w_p|)^2 r_e^{-2}
\bigr)/8.
\end{equation}

Now,
assuming, for definiteness, that $w_p\ge 0$,
let us see what happen if the condition
\begin{equation}
  \label{eq:dw-case}
  \epsilon<\epsilon_c
\end{equation}
is fulfilled.
It can be shown that
$P_{+}(T_{\mn})=P_{+}(z_{\mx})>0$ and 
$P_{-}(T_{\mn})=P_{-}(z_{\mx})<0$.
Referring to Fig.~\ref{fig:crit-epsilon-eps}a,
$P_{+}$ changes its sign 
as $T_{w}$ passes through the point $T_1^{(u)}$
and is negative on the interval $(T_1^{(u)}, T_{2}^{(u)})$.
The endpoints of this interval can be found as
the roots of the polynomial $\epsilon z P_{+}(z)$
and determine the region of unstable cycles
\begin{equation}
  \label{eq:dw-unst-cond-sqr}
  P_{\pm}<0\quad
\text{at}
\quad
T_1^{(u)}<T_{w}<T_2^{(u)},
\end{equation}
where
\begin{equation}
  \label{eq:dw-T12-sqr}
  T_{i}^{(u)}=-2\epsilon^{-1} \ln z_i,
\quad
z_{1,\,2}=(r_e\pm\sqrt{D_{+}})/a_{+}
\end{equation}
and $P_{+}(z_i)=0$.

Outside the interval~\eqref{eq:dw-unst-cond-sqr},
$P_{+}$ is positive and the root of $\epsilon z P_{-}(z)$,
$P_{-}(z_c)=0$, gives the point $T_{w}=T_c^{(s)}$ where
the sign of $P_{-}$ is reversed. 
So, in addition to the unstable cycles, 
we have the low frequency region of stable cycles
\begin{equation}
  \label{eq:dw-stab-cond-sqr}
  P_{\pm}>0\quad
\text{at}
\quad
T_{w}>T_c^{(s)},
\end{equation}
where
\begin{equation}
  \label{eq:dw-Tc-sqr}
  T_{c}^{(s)}=-2\epsilon^{-1} \ln z_c,
\quad
z_{c}=(r_e-\sqrt{D_{-}})/a_{-}.
\end{equation}
The asymptotic formulas for the critical period 
of stable cycles $T_c^{(s)}$
and the limiting periods of unstable cycles
$T_{1,\,2}^{(u)}$ are
\begin{align}
&
  \label{eq:dw-Tc-approx-sqr}
  T_{c}^{(s)}\approx
2\epsilon^{-1} r_e^{-1}
\bigl(
1+|w_p|+2\epsilon+
\sqrt{(1+|w_p|)^2+4\epsilon |w_p|}
\bigr),
\\
&
 \label{eq:dw-Ti-approx-sqr}
  T_{1,\,2}^{(u)}\approx
2\epsilon^{-1} r_e^{-1}
\bigl(
1-|w_p|+2\epsilon\mp
\sqrt{(1-|w_p|)^2-4\epsilon |w_p|}
\bigr).
\end{align}
The corresponding frequencies,
$f_c=1/T_{c}^{(s)}$ and $f_i=1/T_{i}^{(u)}$
might be called the \textit{critical frequency of stable cycles}
and the \textit{limiting frequencies of unstable cycles},
respectively.
These frequencies define two regions:
(a)~the low frequency region of stable cycles with $f_{w}=1/T_{w}<f_c$
and (b)~the high frequency region of unstable cycles
with $f_2<f_{w}<f_1$.

The limiting frequencies, $f_1$ and $f_2$, merge together
at $\epsilon=\epsilon_c$ and the branch of unstable cycles
disappear as the parameter $\epsilon$ further increases.
Fig.~\ref{fig:crit-epsilon-eps}b shows that
the critical value of $\epsilon$, $\epsilon_c$, increases
with the voltage parameter $r_e$.
By contrast, as it can be seen from Eq.~\eqref{eq:dw-epsc-approx-sqr}, 
the polar anchoring reduces the value of $\epsilon_c$
thus suppressing the unstable cycles.

So, there are no unstable cycles
if the polar anchoring parameter is sufficiently large. 
The inhibition of unstable cycles by the polar anchoring 
is clearly evident from 
the curves presented in Fig.~\ref{fig:crit-unst-eps}.
 
Similar to the limiting frequency $f_1$, the critical frequency $f_c$
declines as the parameter $w_p$ increases (see Fig.~\ref{fig:crit-stbl-eps}a).
But the frequency $f_c$ does not vanish and  
the stable cycles are not suppressed at $|w_p|=1$.

In the limit of zero polar anchoring with $w_p=0$,
the limiting frequencies, $f_1$ and $f_2$,
are equal to the threshold frequency
$f_{\mn}=1/T_{\mn}$, $f_1=f_{\mn}$, and the critical frequency
$f_c$, correspondingly. 
The result is that the stable and unstable cycles are separated by 
the branching point $f_2=f_c$.

Splitting of the branching point
occurs as the effect of the polar anchoring induced asymmetry  
when $w_p\ne 0$ and $f_2>f_c$. 
It means that there is the gap separating
the frequency intervals for stable and unstable cycles. 
The width of this gap, which is the frequency difference $f_2-f_c$,
grows with the polar anchoring parameter.
Interestingly, similar effects take place 
for wave numbers of helical structures in asymmetric cholesteric
liquid crystal cells~\cite{Kis:pre:2005}.  

From the above discussion it follows that,
when the parameter $\epsilon$ increases, the limiting frequencies
move towards each other until they coalesce at the critical point 
$\epsilon=\epsilon_c$.
By contrast, the critical frequency $f_c$ is an increasing function
of $\epsilon$ (see Fig.~\ref{fig:crit-stbl-eps}b) and
the frequency range for the stable cycles
expands with the parameter $\epsilon$. 

In the opposite case where $\epsilon\to 0$,
the frequencies $f_c$ and $f_2$ are both vanishing
along with the branch of stable cycles.
In accord with the results of Sec.~\ref{subsec:square-wave-voltage}, 
the limiting frequency $f_1$ approaches 
the critical frequency $1/T_{c}$ (see Eq.~\eqref{eq:T-c-sqr})
for the model~\eqref{eq:dyn-u-approx}.

The parameters $q_{\pm}^{(\st)}=u_{\pm}^{(\st)}-1$
characterizing the up and down states of the cycles
can be computed by substituting Eq.~\eqref{eq:dw-q1-sqr}
and Eq.~\eqref{eq:dw-q0-sqr}
into the relation~\eqref{eq:q-st-sqr}.
For the switching time parameters, the first term on the right hand
side of Eq.~\eqref{eq:tau-pm-sqr} needs to be changed
by using the function~\eqref{eq:dw-G0-sqr}.
The result is
\begin{align}
  \label{eq:dw-tau-pm-sqr}
  \tau_{\pm}=\epsilon^{-1} 
\ln[1+\epsilon\, q_{\mp}^{(\st)}/(r_e\pm w_p-1)]+\delta\tau_{\pm},
\end{align}
where $\delta\tau_{\pm}$ is given by Eq.~\eqref{eq:dlt-tau-sqr}.
It can be shown that the asymptotic expressions for $\tau_{\pm}$
\begin{align}
  \label{eq:dw-tau-pm-approx-sqr}
  \epsilon \tau_{\pm}\approx
\ln[2/(1+z)]+r_e^{-1}
\bigl(
(1+\epsilon) (1-z)/(1+z)\mp w_p
\bigr),
\end{align}
where $z$ is defined in Eq.~\eqref{eq:dw-Ppm-sqr},
are applicable only for stable cycles at sufficiently low frequencies. 

Dependence of the parameter $u_{\pm}^{(\st)}/T_{w}$ 
on the frequency is shown in 
Fig.~\ref{fig:u-st-eps}a and Fig.~\ref{fig:u-st-eps}b
for stable and unstable cycles, respectively.
Similar results for the switching time parameters
$\tau_{\pm}/T_{w}$ are given in Fig.~\ref{fig:tau-pm-eps}. 

It is seen that the curves for the stable and the unstable branches 
reveal qualitatively different behavior.
In particular, by contrast to the unstable cycles,
for the stable cycles,
the polarization of the up state is higher than
that of the down state
and switching up is faster than switching down.
This effect can be loosely described as 
the cycle is shifted towards either stable or metastable
state depending on the cycle stability.

Our concluding remark concerns the sine-wave and triangular voltages.
These cases can be studied in just the same way
as for the model~\eqref{eq:dyn-u-approx}
using the procedure described in
Sec.~\ref{subsec:sine-wave}.
A comparison between the time derivatives of the modified 
functions~\eqref{eq:dw-G-sqr}
\begin{align}
&
  \label{eq:dw-der-G0}
  \pdr{G_{0}(\tau)}{\tau}=(w_p-1+r(\tau))\exp[\epsilon\tau],
\\
&
\label{eq:dw-der-Gplus}
\pdr{G_{+}(\tau)}{\tau}=-(w_p+1+r(\tau))
\exp[\epsilon (\tau-T_{w}/2)]   
\end{align}
and Eq.~\eqref{eq:der-G0-plus} shows that the change of the analytical
expressions does not affect behavior of the functions and the general
results of Sec.~\ref{subsec:sine-wave} are equally applicable to the
model with the double-well potential~\eqref{eq:dw-poten-gen}.  After
performing numerical analysis for these waveforms, we have found that
the predictions of the model discussed in this appendix have not been
changed qualitatively.

%\bibliographystyle{apsrev}
%\bibliographystyle{lc}
%\bibliography{polymer,scatter,lc,quant,hk,flc}

\end{document}